\documentclass[11pt,citesort]{article}
\usepackage{epsfig}
\usepackage{amsfonts}
\usepackage{amsmath}
\usepackage{amssymb}
\usepackage{color}
\usepackage{url}
\usepackage{cite}
\newcommand{\w}{\omega}

\newcommand{\al}{\alpha}
\newcommand{\bt}{\beta}

\newcommand{\simu}{\sigma^{\mu\nu}}

\newcommand{\Nb}{\bar N}
\newcommand{\Fp}{F_\pi}

\newcommand{\tb}{\bar \theta}

\newcommand{\mpi}{m_{\pi}}

\newcommand{\ga}{\gamma}

\newcommand{\dslash}[1]{#1 \llap{/\kern-0.5pt}}
\newcommand{\Dslash}[1]{#1 \llap{/\kern+1.2pt}}
\newcommand{\DDslash}[1]{#1 \llap{/\kern+2.3pt}}
\newcommand{\dslashh}[1]{#1 \llap{/\kern+1pt}}

\newcommand{\bea}{\begin{eqnarray}}
\newcommand{\eea}{\end{eqnarray}}
\newcommand{\bma}{\begin{pmatrix}}
\newcommand{\ema}{\end{pmatrix}}
\newcommand{\nn}{\nonumber}

\begin{document}
\begin{titlepage}

\vspace{2.0cm}

\begin{center}
{\Large\bf
Violations of discrete space-time symmetries\\ 
\vspace{2mm} in chiral effective field theory \vspace{1.3cm}}

{\large \bf   J. de Vries$^{1}$ and Ulf-G. Mei{\ss}ner$^{2,1,3,4}$.}

\vspace{0.5cm}

{\large
$^1$
{\it Institute for Advanced Simulation, Institut f\"ur Kernphysik,
and J\"ulich Center for Hadron Physics, Forschungszentrum J\"ulich,
D-52425 J\"ulich, Germany}}

\vspace{0.25cm}
{\large
$^2$
{\it Helmholtz-Institut f\"ur Strahlen- und Kernphysik and Bethe Center for
Theoretical Physics, Universit\"at Bonn, D-53115 Bonn, Germany}}

\vspace{0.25cm}
{\large
$^3$
{\it JARA\,\,-\,\,Forces and Matter Experiments, Forschungszentrum J\"ulich,
D-52425 J\"ulich, Germany}}

\vspace{0.25cm}
{\large
$^4$
{\it JARA\,\,-\,\,High Performance Computing, Forschungszentrum J\"ulich,
D-52425 J\"ulich, Germany}}

\end{center}

\vspace{1.5cm}

\begin{abstract}
We review recent progress in the theoretical description of the violation of discrete space-time symmetries in hadronic and nuclear systems. 
We focus on parity-violating and time-reversal-conserving interactions which are induced by the Standard Model weak interaction, and on parity- and time-reversal-violating interactions which can be caused by a nonzero QCD $\tb$ term or by beyond-the-Standard Model physics. We discuss the origins of such interactions and review the development of the chiral effective field theory extension that includes discrete symmetry violations. We discuss the construction of symmetry-violating chiral Lagrangians and nucleon-nucleon potentials and their applications in few-body systems.

\end{abstract}

\vfill
\end{titlepage}

\section{Introduction}

Observations of the violation of the discrete space-time symmetries parity (P), charge conjugation (C), and time reversal (T), have played an important role in the construction of the Standard Model (SM). The observation of P violation (PV) in nuclear $\beta$ decay gave rise to the V-A structure of the weak interaction, while observations of CP violation (CPV) lead to the prediction of a third generation of quarks. The pattern of discrete symmetry violations is nowadays accurately described by the SM. Nevertheless, a number of outstanding issues remain.  

Although flavor-conserving PVTC\footnote{We assume the CPT theorem to hold which implies that CP conservation is equal to T conservation (TC).} in the SM is well understood at the level of elementary fields, its manifestation at the nuclear level is not clear. The SM predicts PVTC interactions between quarks, but their appearance at the hadronic level is masked by the nonperturbative nature of low-energy QCD \cite{RamseyMusolf:2006dz, Haxton:2013aca, Schindler:2013yua}.  In fact, a consistent framework that satisfactorily describes all existing PVTC data does not exist at the moment. The problem of how the PVTC four-quark operators appear at the nuclear level thus not only tests the SM weak interaction, but also provides an interesting testing ground for nonperturbative QCD \cite{Schindler:2013yua}.  

So far, all observations of CPV in the K and B sector are explained by the phase of the quark mixing matrix \cite{Kobayashi:1973fv}. However, this phase appears to be insufficient to explain the universal matter-antimatter asymmetry which requires additional CPV \cite{Canetti:2012zc}. The SM contains a second CPV source in the form of the QCD $\tb$ term, which is flavor conserving and severely constrained by the non-observation of the neutron electric dipole moment\footnote{Nonzero EDMs require the violation of both T (and thus CP) and P.} (EDM) \cite{Baker:2006ts}. Because the CKM mechanism predicts extremely small EDMs \cite{Khriplovich:1981ca, Seng:2014lea}, the search for EDMs is a promising method to look for beyond-the-SM (BSM) physics. Current EDM measurements probe high-energy scales comparable to the reach of the Large Hadron Collider. To interpret current EDM limits, and hopefully future nonzero measurements, a solid understanding is needed of how flavor-diagonal P and T violation (PVTV), is manifested at the hadronic and nuclear level \cite{Pospelov:2005pr,Engel:2013lsa, Dekens:2014jka}. 

The third possibility of discrete symmetry breaking consists of P-conserving but T-violating (PCTV) interactions. Such interactions are only induced in the SM via combinations of PVTC and PVTV effects and are therefore very small. Searching for PCTV effects then seems a promising way to look for BSM physics. However, any BSM PCTV interactions can mix with SM PVTC resulting in PVTV interactions that are strongly constrained by EDM experiments  \cite{Khriplovich:1990ef, Engel:1995vv,RamseyMusolf:1999nk}. Direct PCTV measurements involve scattering processes whose precision is outmatched by EDM searches. Nevertheless, scenarios exist where PCTV effects could avoid the EDM constraints \cite{Kurylov:2000ub}. We do not further discuss PCTV, but refer to Refs.~\cite{Song:2011jh,Mereghetti:2013bta,Uzikov:2015aua} for some recent studies.

Traditional approaches of discrete symmetry breaking in nuclear systems are mostly based on one-boson-exchange (OBE) models that describe symmetry breaking in terms of various meson-nucleon interactions. Although these models provide a useful parametrization, it is difficult to link them to the underlying physics and it is unclear whether they capture all aspects of the problem. The understanding of low-energy strong interactions has increased tremendously in the last decades by the use of effective field theories (EFTs), in particular chiral EFT ($\chi$EFT) \cite{Bedaque:2002mn,Epelbaum:2008ga}, the low-energy EFT of QCD. In this work we review the extension of $\chi$EFT that includes the breaking of discrete space-time symmetries.

The $\chi$EFT approach has a number of advantages over OBE models. First of all, there exists a direct link to the microscopic theory at the quark-gluon level. Another advantage is that the symmetry-conserving and -violating interactions among the relevant degrees of freedom, pions and nucleons, are obtained within the same framework which allows for consistent calculations. Such calculations  can be improved order by order in a controlled expansion within $\chi$EFT. Here we discuss the PVTC and PVTV potentials up to next-to-next-to-leading order (N${}^2$LO) where they have a richer form than the corresponding OBE models. Finally, the chiral approach can be extended to multi-nucleon or electromagnetic interactions allowing for a unified treatment of various different observables. Some of these observables are discussed below.

In this work we study low-energy PVTC and PVTV interactions in the framework of $\chi$EFT. In Sec.~\ref{origins} we discuss the structure of symmetry-breaking interactions at the quark-gluon level both from within the SM and beyond. In Sect.~\ref{potentials} we briefly discuss $\chi$EFT and describe the construction of the PCTC, PVTC, and PVTV nucleon-nucleon ($N\!N$) potentials, focusing on their differences and similarities. We also discuss the current status of the low-energy constants (LECs) appearing in the Lagrangians. In Sect.~\ref{observables} we focus on several applications of the obtained potentials. In the PVTC case, we discuss observables in few-body systems and a recent fit of the weak pion-nucleon coupling to existing data. In the PVTV case, we report on calculations of EDMs of light nuclei and their possible implications in the search for BSM physics. We summarize and give an outlook in Sect.~\ref{outlook}. Finally, we refer to several recent reviews on PVTC \cite{Haxton:2013aca,Schindler:2013yua} and PVTV \cite{Engel:2013lsa,Mereghetti:2015rra} where more details on the topics addressed below can be found. 
 
\section{Symmetry violations at the microscopic level}\label{origins}
The various possibilities of discrete symmetry breaking have very different origins. It is important to investigate these origins in order to constrain the operators appearing at lower energies in the $\chi$EFT Lagrangian. In principle, we could start the analysis directly at the hadronic scale by writing down all possible interactions among hadronic degrees of freedom that break the discrete symmetries under investigation.  The disadvantage of such an approach is that valuable information about the hierarchy and structure of the hadronic interactions is lost. The chiral symmetry properties of the microscopic PVTC and PVTV operators constrain the form of the hadronic operators \cite{Kaplan:1992vj,Zhu:2004vw,deVries:2012ab,Mereghetti:2010tp,Bsaisou:2014oka}. It is therefore wise to start the analysis at a scale where QCD is still perturbative, somewhere above the chiral-symmetry-breaking scale $\Lambda_\chi\simeq 1$ GeV. 

\subsection{Origins of PVTC interactions}\label{Opvtc}
Within the SM, PVTC arises from the different gauge-symmetry representations of the chiral fermions. As a consequence, only left-handed quarks and leptons are sensitive to the charged current weak interaction. Both left- and right-handed fermions interact via the neutral current weak interaction, but with different strengths such that P (and C) is still violated. 
Because the physical weak gauge bosons, $W^{\pm}$ and $Z$, have masses much larger than the typical hadronic/nuclear scale, they can be integrated out and matched to effective PVTC four-quark operators. At a scale slightly below $M_W$, the mass of the $W$ boson, the PVTC operators involving the $u$ and $d$ quarks can be conveniently written as
\begin{eqnarray}\label{FQPV}
\mathcal L_{\mathrm{PVTC}} &=& \frac{G_F}{\sqrt{2}} \bigg[ \left(\frac{1}{2}-\frac{1}{3} \sin^2 \theta_W\right)\,V_\mu^a A^{\mu a}  - \frac{1}{3}\sin^2 \theta_W\,I_\mu A^{\mu3}\nonumber\\
&&\phantom{\frac{G_F}{\sqrt{2}} \bigg[}-\sin^2 \theta_W \left(V_\mu^3 A^{\mu 3}-\frac{1}{3}V_\mu^a A^{\mu a} \right)\bigg]+\dots\,\,\,.
\end{eqnarray}
Here we have defined $V_\mu^a = \bar q \gamma_\mu \tau^a q$, $A_\mu^a = \bar q \gamma_\mu \gamma^5 \tau^a q$, and $I_\mu = \bar q \gamma_\mu q$ in terms of the quark doublet $q = (u\,d)^T$. Furthermore, $G_F \simeq 1.16 \cdot 10^{-5}$ GeV${}^{-2}$ is Fermi's constant, $\sin^2\theta_W \simeq 0.23$ defines the weak mixing angle, and we have set the Cabibbo angle to unity. The dots denote operators involving heavier quarks. In particular, operators with strange quarks can have important consequences for nuclear PVTC effects. The Lagrangian in Eq.~\eqref{FQPV} needs to be brought to lower energies via renormalization group (RG) evolution which dresses the coupling constants with $\mathcal O(1)$ QCD factors and induces operators with different color structure \cite{Dai:1991bx,Tiburzi:2012hx}. 

The three operators all break P but have different chiral symmetry properties, with the first transforming as a scalar, the second as an isovector, and the third as an isotensor \cite{Kaplan:1992vj}. Each of the operators thus induces different $\chi$EFT Lagrangians. Although all operators in Eq.~\eqref{FQPV} are proportional to $G_F \sim M_{W}^{-2}$, there is nevertheless a (small) hierarchy in the sizes of the couplings. Due the smallness of $\sin^2 \theta_W\simeq 0.23$, the coupling of the isovector (isotensor) operator is suppressed by a factor $5$ ($2$) with respect to the first operator. This suppression could potentially lead to smaller LECs arising from the isovector operators, but whether this is actually the case depends on, so far, unknown strong matrix elements. In addition, operators involving strange quarks with identical chiral symmetry properties, such as $(\bar s \gamma^\mu s)A_\mu^3$, do not scale with $\sin^2 \theta_W$  and could overcome the suppression factors \cite{Kaplan:1992vj,Meissner:1998pu}.  

\subsection{Origins of PVTV  interactions}\label{Opvtv}
As discussed briefly in the introduction, the SM has two CPV sources. The phase of the quark mass matrix  predominantly manifests in flavor-changing interactions and is responsible for the observed CPV in $K$ and $B$ decays. Via electroweak corrections flavor-diagonal PVTV interactions are induced, but the suppression factors make them immeasurably small with expected experimental accuracies. For instance, the neutron EDM arising from the CKM mechanism lies roughly six orders of magnitude below the current experimental limit \cite{Khriplovich:1981ca, Seng:2014lea}.

The second source of CPV in the SM appears in the strong interaction in the form of the QCD $\tb$ term \cite{'tHooft:1976up}. CPV arising from the $\tb$ term is closely connected to the quark masses and it is useful to consider them simultaneously. Focusing on the lightest two flavors, the non-kinetic part of the QCD Lagrangian can be written as
\begin{equation}
\mathcal L_{m,\tb} = -\left(e^{i \rho}\,\bar q_L M q_R + e^{-i \rho}\,\bar q_R M q_L \right)- \frac{\theta g_s^2}{64\pi^2}\epsilon^{\mu\nu\alpha\beta}G_{\mu\nu}^a G_{\alpha\beta}^a\,\,\,, 
\end{equation}
in terms of the gluon field strength $G$, the diagonal real mass matrix $M = \bar m (1+\varepsilon \tau^3)$ where $\bar m=(m_u+m_d)/2$ and $ \varepsilon=(m_u-m_d)/(m_u+m_d)$, the strong coupling constant $g_s$, the overal phase $\rho$, and the angle $\theta$. The interaction proportional to $\theta$ is a total derivative but contributes to the action via instantons \cite{'tHooft:1976up}. For $\chi$EFT purposes it is useful to perform an anomalous $U(1)_A$ transformation to eliminate the $\theta$ term in favor of a complex mass term proportional to $\tb = \theta+2\rho$. A subsequent non-anomalous $SU(2)_A$ rotation can be used to cast the complex mass term in isoscalar form \cite{Baluni:1978rf, Mereghetti:2010tp} 
\begin{equation}\label{theta}
\mathcal L_{m,\tb} = -\bar m \left( \bar q q+ \varepsilon\,\bar q \tau^3 q - \frac{1-\varepsilon^2}{2}\tb\,\bar q i\gamma^5 q\right)\,\,\,,
\end{equation}
where we used $\tb \ll 1$ (the current constraint on the neutron EDM forces $\tb < 10^{-10}$ \cite{Crewther:1979pi}). The lack of explanation for this extreme smallness is usually called the strong CP problem. A popular solution is the so-called Peccei-Quinn mechanism \cite{Peccei:1977hh} which dynamically explains the smallness of $\tb$ at the cost of a so far unmeasured new particle, the axion. 
 
Generic BSM models contain additional CPV phases that can induce larger EDMs than the SM \cite{Pospelov:2005pr}. 
Since many BSM variants exist, from the point of view of low-energy experiments it is preferable to perform a model-independent analysis. Considering the success of the SM, it is likely that any BSM physics appears at a scale considerably higher than the electroweak scale. This scale separation makes it possible to treat the SM as the dimension-four and lower part of a more general EFT containing higher-dimensional operators. Such operators must conserve the all-important SM Lorentz and $U(1)_Y\times SU(2)_L\times SU(3)_c$ gauge symmetries that put strong constraints on their form \cite{Grzadkowski:2010es}.
For our purposes, the first relevant operators appear at dimension six and are suppressed by two powers of the scale, $\Lambda_{CPV}$, where the additional CPV is assumed to appear. BSM CPV can be studied in a model-independent way by adding all CPV dimension-six operators \cite{deVries:2011an,deVries:2012ab}. The great advantage is that it is not necessary to choose a specific SM extension, but, if so wanted, the approach can easily be matched to specific models \cite{Dekens:2014jka}.

The advantage of starting the EFT analysis at $\Lambda_{CPV}$ is that we make full use of the SM gauge symmetries. The disadvantage is that we must evolve the operators to $\Lambda_\chi$ which is more involved than in the PVTC case.  The RG evolution involves QCD and electroweak corrections \cite{Degrassi:2005zd, Hisano:2012cc, Dekens:2013zca} and heavy SM particles need to be integrated out at their thresholds. A full study has been performed for operators involving the lightest two quarks \cite{Dekens:2013zca}, while the evolution of operators with heavier quarks is more fragmented, see for instance \cite{Kamenik:2011dk,Sala:2013osa, Chien:2015xha}. QCD RG typically mildly suppresses the operators when evolved to lower scales and mixes several operators, making it more difficult to identify the high-energy origin from EDM experiments.

Despite these difficulties, the number and form of PVTV interactions around $\Lambda_\chi$ is very limited \cite{deVries:2012ab}. Being interested in hadronic and nuclear PVTV we neglect (semi-)leptonic operators. The first operators that appear are then the quark\footnote{Here we focus on the lightest two quark flavors and neglect strange q(C)EDMs.} EDMs and chromo-EDMs (CEDMs)
\bea \label{dim5}
\mathcal L_{\mathrm{PVTV}} = -\frac{1}{2} \, \bar q (d_0 + d_3 \tau^3)i \simu \ga_5  q\, F_{\mu\nu} -\frac{1}{2}\ \bar q (\tilde d_0 + \tilde d_3 \tau^3) i\simu \ga_5 t_a q\, G^a_{\mu\nu}\,\,\,,
\eea
where $F_{\mu\nu}$ denotes the photon field strength and $t_a$ the generators of $SU(3)$.  
Although these operators appear to be dimension five, $SU(2)_L$ gauge invariance forces them to be coupled to the Higgs field at high energies making them effectively dimension six. After electroweak symmetry breaking, the Higgs field takes on its vacuum expectation value (vev) resulting in a quark mass (the actual value of the mass is model dependent and could be larger or smaller) appearing in the scalings $d_{0,3},\tilde d_{0,3}\sim \bar m \Lambda_{CPV}^{-2}$.  

The next operator is the Weinberg operator, also called the gluon CEDM (gCEDM) \cite{Weinberg:1989dx,Braaten:1990zt}
\bea \label{dW}
\mathcal L_{\mathrm{PVTV}} = d_W \frac{1}{6}f_{abc}\varepsilon^{\mu\nu\al\bt}G^a_{\al\bt}G^b_{\mu\rho}G_{\nu}^{c \, \rho}\,\,\,,
\eea
which is the only purely gluonic CPV operator at dimension six. 
Finally, there are several PVTV four-quark operators
 \bea \label{FQ}
\mathcal L_{\mathrm{PVTV}} = \mathrm{Im}\,\Sigma\,(\bar q q\,\bar q i\gamma^5 q - S^a P^a) +  \mathrm{Im}\,\Xi\, \epsilon^{3ab} (S^a P^b)+\dots\,\,\,,
\eea
in terms of $S^a = \bar q \tau^a q$ and $P^a = \bar q \tau^a i\gamma^5 q$,  and the dots denote two similar operators with different color structure \cite{Dekens:2013zca}. The coupling constants $d_W$, $\mathrm{Im}\,\Sigma$, and $\mathrm{Im}\,\Xi$ parametrize BSM physics and scale as $\Lambda_{\mathrm{CPV}}^{-2}$. The operators have very different origins. The operators proportional to $d_W$ and $\mathrm{Im}\,\Sigma$ conserve all gauge symmetries and appear directly at $\Lambda_{\mathrm{CPV}}$ \cite{RamseyMusolf:2006vr}. Because they are invariant under chiral symmetry transformations their low-energy consequences are similar and in what follows we only focus on the gCEDM. 
The operator proportional to $\mathrm{Im}\,\Xi$ is induced after electroweak symmetry breaking via a $W^\pm$ exchange\footnote{This exchange involves a left- and right-handed $W^\pm$ current and the resulting operator is usually called the four-quark left-right operator (FQLR).} between quarks \cite{Ng:2011ui}. 


\section{Chiral Lagrangians and nucleon-nucleon potentials}\label{potentials}
Now that we have identified the structure of PVTC and PVTV interactions slightly above $\Lambda_\chi$, we must understand how these interactions manifest themselves at lower energies where QCD becomes nonperturbative. To do so, we use the framework of $\chi$EFT. By constructing the most general Lagrangian in terms of the relevant low-degrees of freedom that incorporates the symmetries of the microscopic
theory (QCD), we obtain an EFT, called chiral perturbation theory ($\chi$PT), which is the low-energy equivalent of QCD.
The big advantage of $\chi$PT is that observables can be calculated in perturbation theory. The expansion parameter is  $p/\Lambda_\chi$, where $p$ is the momentum scale appearing in the process. The nonperturbative nature of low-energy QCD is captured by the low-energy constants (LECs) associated with each interaction. These LECs need to be fitted to data or calculated with nonperturbative methods such as lattice QCD.  $\chi$PT was originally formulated for mesons, but has been extended to the nucleon and multi-nucleon sectors, where it is usually called $\chi$EFT. A big success of $\chi$EFT is the derivation of the structure and hierarchy of multi-nucleon interactions. The strong $N\!N$ potential has been derived up to N$^4$LO \cite{Epelbaum:2014sza} and successfully describes many few-body processes. Reviews on $\chi$PT and $\chi$EFT can be found in Refs.~\cite{Bernard:2006gx,Bedaque:2002mn,Epelbaum:2008ga}.

A special role in $\chi$EFT is played by the pion triplet. Pions emerge as Goldstone bosons of the spontaneously broken chiral symmetry to its isospin subgroup $SU(2)_L\times SU(2)_R\rightarrow SU(2)_I$. Interactions of Goldstone bosons are proportional to their momenta which explains the perturbative nature of $\chi$PT at low energies.
Because chiral symmetry is only an approximate symmetry of QCD, being broken by the quark masses and charges, the pions obtain a small mass and become pseudo-Goldstone bosons. However, the smallness of the symmetry-breaking terms ensures that they can be incorporated in the expansion. The $\chi$EFT Lagrangian is then obtained by constructing all chiral-invariant interactions and all interactions that break chiral symmetry in the same way as the chiral-breaking sources at the quark level. In principle, an infinite number of interactions exist, but they can be ordered by the chiral index $\Delta = d + n/2-2$, where $d$ counts the number of derivatives\footnote{Because $m_N/\Lambda_\chi$ is not a small number, derivatives acting on nucleon fields are in principle not suppressed. In this work we resolve the issue by considering heavy-baryon $\chi$PT \cite{Jenkins:1990jv,Bernard:1992qa}, where the nucleon is treated nonrelativistically and the large nucleon mass $m_N$ is removed from the nucleon propagators.} and quark mass insertions (as $ m_q\sim\mpi^2\sim p^2$, each quark mass insertion increases $d$ by $2$) and $n$ the number of nucleon fields \cite{Weinberg:1978kz,Gasser:1983yg}.  We use this definition of the chiral index throughout this review (Weinberg power counting), but  refer to Ref.~\cite{Valderrama:2014vra} for discussions of possible enhancements (i.e. lower chiral indices) for short-range nuclear current operators.
Because the PVTC and PVTV interactions are associated with very small parameters, \textit{e.g.} $G_F \Lambda_\chi^2 \simeq 10^{-5}$ and $\tb <10^{-10}$, they can be included in the expansion as well. In fact, the interactions are so weak that only $\chi$EFT operators  linear in the symmetry-breaking parameters must be considered. 

An important ingredient in few-body calculations is the $N\!N$ potential which can be derived from the $\chi$EFT Lagrangians \cite{Weinberg:1990rz,Ordonez:1993tn}. Weinberg showed \cite{Weinberg:1990rz} that the importance of a connected irreducible diagram with $N$ nucleons, $L$ loops, and $V_i$ insertions of an interaction with chiral index $\Delta_i$ is given by the index
\begin{equation}
\nu = -4 + 2N + 2L +\sum_i V_i \Delta_i\,\,\,.
\end{equation}
The index $\nu$ is bounded from below which allows for an order-by-order derivation of the PCTC, PVTC, and PVTV $N\!N$ potentials.

\subsection{PCTC chiral interactions}\label{VPCTC}
 We begin by listing the most important interactions that conserve $P$, $T$, and $C$ and originate in the QCD Lagrangian. The construction of the chiral operators is well known and explained in various reviews \cite{Bernard:2006gx,Epelbaum:2008ga} and here we only discuss some of the relevant operators. Because chiral symmetry is realized in nonlinear fashion in $\chi$EFT, each interaction is associated with other interactions involving more pions. For simplicity, we only focus here on the operators with the least number of pions. The operators with the lowest chiral index $\Delta =0$  are given by
 \begin{eqnarray}\label{PCTC0}
 \mathcal L^{(0)}_{\mathrm{PCTC}} &=& \frac{1}{2} \partial_\mu \vec \pi \cdot \partial^\mu \vec \pi  -\frac{1}{2}\mpi^2\vec \pi^2 + \Nb i v\cdot \partial N -\frac{g_A}{F_\pi}\partial_\mu \vec \pi \cdot \Nb \vec \tau S^\mu N\nonumber\\
 &&-\frac{1}{2}C_S(\Nb N)(\Nb N)+ 2 C_T (\Nb S_\mu N)(\Nb S^\mu N)+\dots\,\,\,,
 \end{eqnarray}
 in terms of the pion triplet $\vec \pi$, the nucleon doublet $N= (p\, n)^T$, the nucleon velocity $v^\mu$ and spin $S^\mu$, the pion decay constants $F_
 \pi =92.4$ MeV, and three LECs $g_A$, $C_S$, and $C_T$. The dots denote associated interactions with more pions such as the Weinberg-Tomozawa vertex. Apart from the pion mass term, which involves pions without a derivative, all operators originate in the chiral-invariant part of the QCD Lagrangian.

\begin{figure}[!t]
\centering
\includegraphics[width=0.95\textwidth]{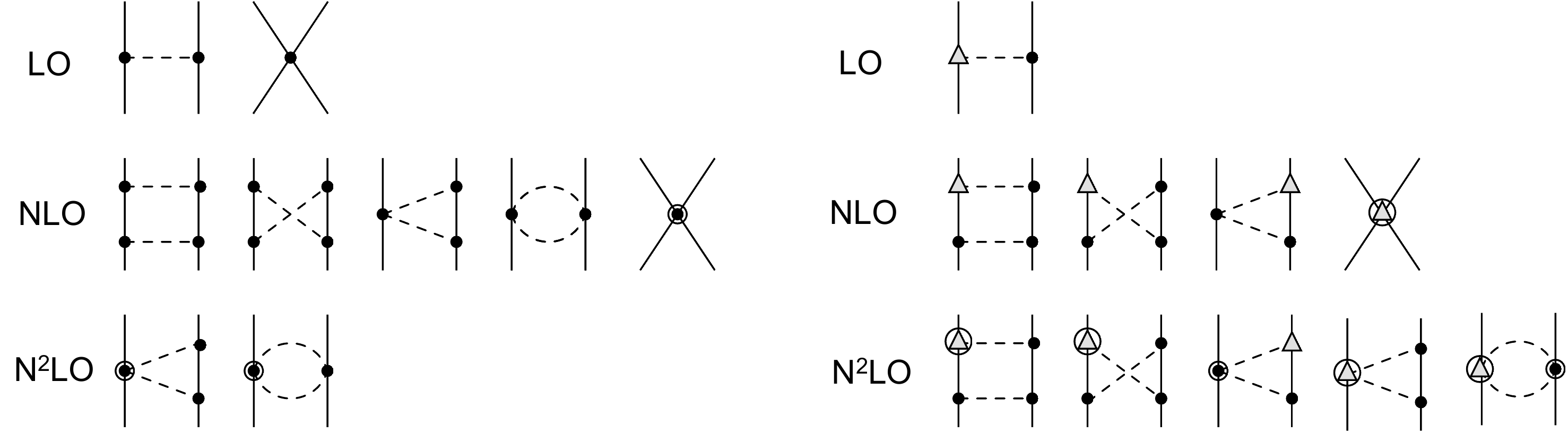} 
\caption{\small Chiral expansion of the PCTC (left) and PVTC (right) nucleon-nucleon force up to N${}^2$LO. Solid and dashed lines denote nucleons and pions, respectively. Circles and triangles represent, respectively, LO PCTC and PVTC interactions while circled vertices represent NLO interactions. Corrections to the one-pion-exchange potential and three-body forces are not shown.
}
 \label{FigPot1}
\end{figure}

What is important to notice is that chiral symmetry and P and T conservation ensure that the leading-order (LO) pion-nucleon vertex comes with a derivative, while the LO $N\!N$ interactions do not. Therefore, both interactions have chiral index $\Delta =0$ and contribute to the LO $N\!N$ potential (with $\nu=0)$. At NLO, with chiral $\nu=2$ and thus a relative suppression of $(p/\Lambda_\chi)^2$, there are additional contact terms with two derivatives, and two-pion-exchange (TPE)  diagrams involving LO vertices. At N${}^2$LO additional TPE diagrams appear which involve $\pi\pi$-nucleon interactions with chiral index $\Delta=1$, the so-called $c_i$ interactions \cite{Bernard:1995dp}, which also contribute to three-body forces appearing at the same order.  The schematic hierarchy of the PCTC potential is shown in the left panel of Fig.~\ref{FigPot1}. 
We do not discuss higher-order corrections to the Lagrangian or potential, but refer to Refs.~\cite{Epelbaum:2004fk,Epelbaum:2014sza} for the full expressions.

\subsection{PVTC chiral interactions}\label{PVTCpot}
The four-quark interactions in Eq.~\eqref{FQPV} give rise to PVTC hadronic interactions. The three four-quark interactions transform, respectively, as a chiral singlet, an isovector ($\Delta I=1$), and an isotensor ($\Delta I=2$) \cite{Kaplan:1992vj}. This implies that the first operator can only induce pionic operators with derivatives, while the other two can lead to non-derivative pionic interactions. Only the isovector interaction leads to an interaction with chiral index\footnote{Although, this PVTC interaction has a lower chiral index than the LO PCTC interactions in Eq.~\eqref{PCTC0}, this of course does not imply that the interaction is larger. All PVTC interactions are proportional to $G_F$ which greatly suppresses their effects. The chiral index should only be used to probe the relative importance with respect to interactions with the same transformations under $C$, $P$, and $T$.} 
$\Delta =-1$:
\begin{equation}
\mathcal L_{\mathrm{PVTC}}^{(-1)} = \frac{h_\pi}{\sqrt{2}}\Nb(\vec \pi \times \vec\tau)^3 N\,\,\,,
\end{equation}
where $h_\pi$ is the (in)famous weak pion-nucleon coupling constant whose size is still unknown, despite significant theoretical and experimental effort. In contrast to the PCTC case, there appear no $N\!N$ contact interactions at LO. Such terms require at least one derivative and have chiral index $\Delta =1$. Thus, the LO PVTC potential consists only of a one-pion-exchange (OPE) contribution:
\begin{equation}
V^{(-1)}_{\text{PVTC}}
= - \frac{g_{A}h_\pi}{ 2\sqrt{2} F_\pi} i(\vec \tau_1\times \vec \tau_2)^3 \frac{(\vec \sigma_1+\vec \sigma_2)\cdot \vec q }{\mpi^2+q^2}\,\,\,,
\label{onepion}
\end{equation}
in terms of the nucleon spin $\vec \sigma_{1,2}$ and the momentum transfer flowing from nucleon $(1)$ to nucleon $(2)$: $\vec q = \vec p - \vec p^{\,\prime}$ ($q = |\vec q\,|$), where $\vec p$ and $\vec
p^{\,\prime}$ are the relative momenta of the incoming and
outgoing nucleon pair in the center-of-mass frame. 

Because the LO potential consists of a single term, it might be expected that this term dominates hadronic and nuclear PVTC. From the existing data it should then be possible to fix the size of $h_\pi$ from which other processes can be predicted. Unfortunately, the situation is more complicated for two main reasons:
\begin{itemize}
\item The LO potential changes the total isospin of the interacting nucleon pair and therefore does not contribute to PVTC effects in proton-proton ($pp$) scattering. As a significant part of the nonzero PVTC measurements has been made in this process, higher-order corrections are required to analyze the data.
\item The division of the potential into LO, NLO, .... , is based on an expansion in $p/\Lambda_\chi$. However, as discussed in Sec.~\ref{Opvtc}, the isovector four-quark operator is suppressed by a factor $\sin^2 \theta_W \sim 1/5$. Large $N_c$ arguments indicate that $h_\pi$ could even be further suppressed \cite{Kaiser:1989ah,Meissner:1998pu, Phillips:2014kna}. Thus, formally higher-order corrections might be larger than expected because of dimensionless factors not captured by the chiral counting.
\end{itemize}

A great advantage of $\chi$EFT is that higher-order corrections can be systematically calculated. The NLO potential (with $\nu=1$) was first obtained in Ref.~\cite{Zhu:2004vw} and shown to consist\footnote{NLO corrections to the OPE potential can be absorbed in redefinitions of the LO and NLO LECs \cite{Zhu:2000fc, Viviani:2014zha}.} of TPE diagrams proportional to $h_\pi$ and $N\!N$ contact interactions \cite{Girlanda:2008ts}. 
The TPE contributions suffer from ultraviolet divergences that are absorbed, together with the associated scale dependence, by the contact terms. We show results using spectral function regularization with cut-off $\Lambda_S$ (varied between $500$ and $700$ MeV in the discussions below) which was introduced in Ref.~\cite{Epelbaum:2003gr} to improve the convergence of the PCTC potential. In this way, the PVTC and PCTC potential are regularized in the same way. By taking the spectral cut-off $\Lambda_S\rightarrow \infty$, the results in dimensional regularization are retrieved. The TPE contributions are then given by \cite{Zhu:2004vw,Kaiser:2007zzb,deVries:2014vqa}
\begin{eqnarray}\label{NLOTPE}
V^{(1)}_{\text{PVTC}}
(\Lambda_S)
&=&   -\frac{ g_A h_\pi}{2 \sqrt{2}\Fp} \frac{1}{(4\pi \Fp)^2}\left[i (\vec \tau_1\times \vec \tau_2)^3 (\vec \sigma_1+\vec \sigma_2)\cdot \vec q\right] \left(g_A^2 \frac{8\mpi^2+3q^2}{\omega^2} - 1 \right) L(q,\Lambda_S)\nonumber\\
&& +  \frac{ g_A^3 h_\pi}{2 \sqrt{2}\Fp}  \frac{4}{(4\pi \Fp)^2}\left[i (\vec \tau_1+ \vec \tau_2)^3 (\vec \sigma_1\times \vec \sigma_2)\cdot \vec q\right] L(q, \Lambda_S)\,\,\,,
\end{eqnarray}
in terms of the loop functions
\begin{equation}
\w = \sqrt{q^2+4\mpi^2}\,\,, \hspace{3mm} L(q,\Lambda_S)= \frac{\w}{2q} \log\left(\frac{\Lambda_S^2 \w^2 +q^2 s^2 + 2 \Lambda_S s \w q}{4\mpi^2(\Lambda_S^2+q^2)}\right)\,\,,\hspace{3mm} s = \sqrt{\Lambda_S^2-4\mpi^2}\,\,\,.
\end{equation}
The first term in Eq.~\eqref{NLOTPE} has the same spin-isospin structure as the OPE potential and is therefore not very interesting. However, the second term induces ${}^1S_0\leftrightarrow {}^3 P_0$ transitions and give the first contributions to $pp$ scattering. 

At the same order as the TPE diagrams, we find five\footnote{This number can be understood by noticing that there are five possible $S\leftrightarrow P$ couplings. One ${}^3S_1 \leftrightarrow {}^1 P_1$ transition, one ${}^3S_1 \leftrightarrow {}^3 P_1$ transition, and three, one for each value of $m_t$, ${}^1S_0 \leftrightarrow {}^3 P_0$ transitions. The isospin properties of the PVTC four-quark operators are rich enough to induce all five $N\!N$ operators at the same order.}
 $N\!N$ contact interactions. These can be written in various ways \cite{Girlanda:2008ts, Phillips:2008hn}, and here we use the following parametrization \cite{deVries:2014vqa}
\begin{eqnarray}\label{contactPV}
V^{(1)}_{\text{PVTC}}
&=&  \frac{C_0}{ \Fp \Lambda_\chi^2}  (\vec \sigma_1 - \vec \sigma_2)\cdot (\vec p + \vec p^{\,\prime} )\nonumber\\
&& + \frac{1}{ \Fp \Lambda_\chi^2} \left(C_1 + C_2\frac{(\vec\tau_1+\vec\tau_2)^3}{2} + C_3\frac{\vec \tau_1\cdot \vec \tau_2-3 \tau_1^3 \tau_2^3 }{2} \right) i(\vec \sigma_1 \times \vec \sigma_2) \cdot \vec q\nonumber\\
&& + \frac{C_4}{ \Fp \Lambda_\chi^2} i(\vec \tau_1\times \vec \tau_2)^3(\vec \sigma_1 + \vec \sigma_2)\cdot \vec q\,\,\,.
\end{eqnarray}
All together, the NLO PV potential depends on six LECs which need to be fitted to experiments or calculated with nonperturbative techniques.

At N${}^2$LO ($\nu=2$) several additional TPE diagrams appear, which have been calculated in Ref.~\cite{deVries:2014vqa}. The first part involves no new LECs and is proportional to $\pi h_\pi\,c_4$ and has the same spin-isospin properties as the second term in Eq.~\eqref{NLOTPE}. Because of the large size of $c_4 \simeq 3.4$ GeV${}^{-1}$ \cite{Epelbaum:2004fk}, explained by underlying $\Delta$ and $\rho$-meson resonances, and the enhancement by a factor $\pi$, this term can be expected to dominate the N${}^2$LO potential unless $h_\pi$ is very small. The second part depends on five new PVTC pion-nucleon and pion-pion-nucleon LECs \cite{Kaplan:1992vj}, which will be difficult to fit to the scarce data. At this order we encounter the first contributions to PVTC three-body forces which have not been studied\footnote{PVTC three-body forces in pionless EFT were studied in Ref.~\cite{Griesshammer:2010nd}.} and partially depend on new LECs. The hierarchy of the PVTC potential is sketched in the right panel of Fig.~\ref{FigPot1}.

\subsubsection{Estimates of the PVTC LECs}
\begin{table}[t]
\begin{center}\small
\begin{tabular}{||c|ccc||}
\hline
LEC& DDH `best' value &DDH range& KMW\\
\hline
\rule{0pt}{3ex}
$h_\pi$ & $\phantom{-}0.46\cdot 10^{-6}$ & $\phantom{-}(0.0\rightarrow 1.2)\cdot 10^{-6}$& $\phantom{-}0.10\cdot 10^{-6}$\\
$C_0$ & $\phantom{-0}4.7\cdot 10^{-6}$ & $\,\,(-5.0\rightarrow 13)\cdot 10^{-6}$& $\phantom{-}0.89\cdot 10^{-6}$\\
\rule{0pt}{2ex}
$C_1$ & $\phantom{-0}1.2\cdot 10^{-6}$ & $\,(-2.5\rightarrow 4.5)\cdot 10^{-6}$&$\phantom{-}0.11\cdot 10^{-6}$\\
\rule{0pt}{2ex}
$C_2$ & $\,\,\,-2.2\cdot 10^{-6}$ & $\!\!\!(-5.0\rightarrow -0.2)\cdot 10^{-6}$& $-0.66\cdot 10^{-6}$\\
\rule{0pt}{2ex}
$C_3$ & $\phantom{-0}1.0\cdot 10^{-6}$ &$\,\,\,\,\,\,(0.8\rightarrow 1.2)\cdot 10^{-6}$& $\phantom{-}0.41\cdot 10^{-6}$\\
\rule{0pt}{2ex}
$ C_4$ & $\phantom{-}0.25\cdot 10^{-6}$ & $\,(-0.1\rightarrow 0.7)\cdot 10^{-6}$&$-0.05\cdot 10^{-6}$\\
 \hline

\end{tabular}
\end{center}
\caption{\small Predictions \cite{deVries:2014vqa} of the LECs $C_i$ based on resonance saturation and two different sets of DDH parameters.  The first two columns correspond to, respectively, the best values and range from Ref.~\cite{Desplanques:1979hn}, while the third column is based on the predictions of Refs.~\cite{Kaiser:1989ah,Meissner:1998pu}.}
\label{table1}
\end{table}

Although $\chi$EFT allows the determination of the hierarchy and form of the potential, the LECs cannot be obtained from symmetry considerations alone. The PVTC LECs, in particular $h_\pi$, have been studied with various techniques with varying precision and sophistication. The simplest estimates are obtained from naive-dimensional analysis (NDA) \cite{Manohar:1983md, Weinberg:1989dx} which predicts $h_\pi \sim C_i \sim \mathcal O(G_F F_\pi \Lambda_\chi) \sim 10^{-6}$.
This should be seen as an order-of-magnitude estimate which roughly probes the size of PVTC in nuclear systems.

The most applied approach to hadronic PVTC is the one-meson exchange model (or the DDH model, after the authors of Ref.~\cite{Desplanques:1979hn}), in which PVTC is described by the exchange of a single pion, $\rho$-, or $\omega$-meson. The exchange of a charged pion gives rise to the same potential as Eq.~\eqref{onepion}, while exchanges of the heavier mesons give rise to different structure with shorter range. The DDH potential then depends on $h_\pi$ and six constants associated with heavier mesons. In Ref.~\cite{Desplanques:1979hn}, $h_\pi$ was estimated using the quark model and $SU(6)$ symmetry, finding the reasonable range $0\leq h_\pi \leq 1.2\cdot 10^{-6}$ and a ``best" value of $h_\pi \simeq 4.6 \cdot 10^{-7}$. In similar fashion, a range and best value for the other constants were estimated. 

The DDH parameters were calculated using a soliton description of the nucleon in Refs. \cite{Kaiser:1989ah, Meissner:1998pu}, finding $h_\pi \simeq 1 
\cdot 10^{-7}$. Similar small values of $h_\pi$ were found in a first lattice QCD\footnote{Disconnected diagrams were not included nor was the result extrapolated to the physical pion mass.} calculation \cite{Wasem:2011zz} and a large-$N_c$ analysis \cite{Phillips:2014kna}. The smaller values of $h_\pi \simeq 10^{-7}$ seem to be in better agreement with data. The absence of a PVTC signal in $\gamma$-ray emission from ${}^{18}$F leads to a strong bound on $h_\pi \leq 1.3 \cdot 10^{-7}$ \cite{Adelberger:1983zz,Haxton:1981sf,Page:1987ak}. On the other hand, the measurement of the Cs anapole moment \cite{Wood:1997zq} prefers a larger value $h_\pi \simeq 10^{-6}$ \cite{Flambaum:1997um,Haxton:2001mi}, but suffers from larger nuclear uncertainties. This confusing situation can be cleared up by analyzing PVTC signals in few-body experiments where the nuclear theory is better under control.

By use of resonance saturation, the contact LECs $C_i$ can by estimated in terms of the DDH parameters \cite{Haxton:2013aca}. After integrating out the $\rho$- and $\omega$-meson, the DDH potential (apart from the OPE part) collapses into the five contact interactions in Eq.~\eqref{contactPV} plus terms suppressed by powers of $(p/m_{\rho,\omega})^2$. The estimates of the DDH parameters can then be used to predict the sizes of $C_i$. As the $\chi$EFT potential contains explicit TPE contributions that are absent in the DDH potential, these must be subtracted to get a sensible comparison \cite{Epelbaum:2001fm}. By doing so, Ref.~\cite{deVries:2014vqa} obtained the estimates in Table \ref{table1} for the LECs $C_i$ based on two sets of DDH parameters. The predictions vary by roughly an order of magnitude reflecting the large uncertainty.

\subsection{PVTV chiral interactions}
We now turn to the $\chi$EFT Lagrangians induced by the PVTV sources described in Sect.~\ref{Opvtv}. In the PVTC case, the three four-quark operators in Eq.~\ref{FQPV} were all proportional to $G_F$ and had roughly equal strengths. We therefore could construct one PVTC $\chi$EFT Lagrangian with a clear hierarchy of interactions. Life is not as simple in the PVTV case. As no nonzero flavor-diagonal PVTV measurements have been made, we can only put upper limits on the coupling constants $\tb$, $d_{0,3}$, $\tilde d_{0,3}$, $d_W$, $\mathrm{Im}\,\Sigma$, and $\mathrm{Im}\,\Xi$ and we know nothing about  their relative sizes. By focusing on specific BSM scenarios it is possible to calculate the hierarchy of parameters, but in this way we are no longer working in a model-independent fashion. Consequently, we must construct the PVTV $\chi$EFT Lagrangian for each source separately. This has been described in detail in Refs.~\cite{Mereghetti:2010tp,deVries:2012ab,Bsaisou:2014oka} and here we focus on the most important findings.

Because we construct the $\chi$EFT Lagrangian for each PVTV source, the same chiral interaction can have different chiral indices depending on the PVTV source. To illustrate this we look at the PVTV pion-nucleon interactions. The simultaneous breaking of chiral symmetry, P, and T, allows for three different non-derivative vertices:
\begin{eqnarray}\label{g012}
\mathcal L^{\pi N}_{\mathrm{PVTV}} = \bar g_0 \Nb \vec \tau \cdot \vec \pi N +  \bar g_1 \Nb \pi^3 N +   \bar g_2 \Nb \tau^3 \pi^3 N\,\,\,,
\end{eqnarray}
in terms of  three LECs $\bar g_{0,1,2}$. The $\tb$ term in Eq.~\eqref{theta} breaks chiral symmetry as a complex isoscalar quark mass and can only directly induce $\bar g_0$ which then has chiral index $\Delta_\theta (\bar g_0) = -1$. To generate $\bar g_1$, we require an insertion of the quark mass difference, raising the chiral index to $\Delta_\theta (\bar g_1) = 1$. Even that is not enough to induce $\bar g_2$ which requires another insertion of the quark mass difference  or a photon exchange\footnote{Here we follow Ref.~\cite{Epelbaum:2004fk} and count $\alpha_{\mathrm{em}}/(4\pi)$ as $(p/\Lambda_\chi)^4$.} $\Delta_\theta (\bar g_2) = 3$. On the other hand, the qCEDMs in Eq.~\eqref{dim5} break both chiral and isospin symmetry (if $|\tilde d_0| \simeq |\tilde d_3|$) and we obtain: $\Delta_{\tilde d_q} (\bar g_0) = \Delta_{\tilde d_q} (\bar g_1)=-1$, while a quark mass insertion is needed for $\bar g_2$: $\Delta_{\tilde d_q} (\bar g_2) =1$. Clearly, different PVTV sources induce different $\chi$EFT Lagrangians.

\begin{table}[t]
\begin{center}\small
\begin{tabular}{||c|ccccc||}
\hline
\rule{0pt}{3ex}
& $\tb$ &qCEDM($\tilde d_{0,3}$) & qEDM($d_{0,3}$) & gCEDM ($d_W$) & FQLR ($\mathrm{Im}\,\Xi$)\\
\hline
\rule{0pt}{3ex}
$\bar g_0$ & $-1$ & $-1$ & $3$& $1$ & $1$ \\
\rule{0pt}{3ex}
$\bar g_1$ & $1$ & $-1$ & $3$& $1$ & $-1$\\
\rule{0pt}{3ex}
$\bar g_2$ & $3$ & $1$ & $3$& $2$ & $1$\\
\rule{0pt}{3ex}
$\bar \Delta$ & $0$ & $0$ & $4$& $2$ & $-2$\\
\rule{0pt}{3ex}
$ \bar C_{1,2}$ & $1$ & $1$ & $5$& $1$ & $3$\\
\rule{0pt}{3ex}
$\bar d_{0,1}$ & $1$ & $1$ & $1$& $1$ & $1$\\
 \hline
\end{tabular}
\end{center}
\caption{\small The chiral indices $\Delta = d+n/2-2$ of various PTVT chiral interactions \cite{Mereghetti:2010tp,deVries:2012ab,Bsaisou:2014oka}, where $d$ counts the number of derivatives, photon fields, and quark mass insertions and $n$ the number of nucleon fields appearing in the operator. The chiral indices should only be used to probe the relative strength of the interactions for each PVTV source separately. The  LECs arising from different sources cannot be compared model-independently due the unknown sizes of the parameters $\tb$, $d_{0,3}$, $\tilde d_{0,3}$, $d_W$, and $\mathrm{Im}\,\Xi$. }
\label{table2}
\end{table}

The same game can be played for other PVTV interactions. It turns out that for all sources, the operators with the lowest chiral index are given by\footnote{The simultaneous violation of $P$, $T$, and isospin symmetry allows for the presence of pion tadpoles $\mathcal L = f_{\mathrm{tad}}\pi_3$. These terms can be eliminated via field redefinitions at the price of additional PVTV interactions in other sectors \cite{Mereghetti:2010tp,deVries:2012ab,Bsaisou:2014oka}. The Lagrangian in Eq.~\eqref{Lpvtv} is interpreted as subsequent to these redefinitions.}
\begin{eqnarray}\label{Lpvtv}
 \mathcal L_{\mathrm{PVTV}}
  &=& \bar g_0 \Nb \vec \tau \cdot \vec \pi N +  \bar g_1 \Nb  \pi^3 N+m_N\bar \Delta\,\pi_3\,\vec \pi^2-2 \Nb(\bar d_0 + \bar d_3 \tau^3)S^\mu N v^\nu F_{\mu\nu}\nonumber\\
  &&+\bar C_1\Nb N\,\partial_{\mu}(\Nb S^{\mu}N)+\bar C_2\Nb \vec \tau N\cdot\partial_{\mu}(\Nb \vec \tau S^{\mu}N)\,\, ,
\end{eqnarray}
in terms of seven LECs \cite{deVries:2011an,deVries:2012ab}. Which of these interactions is actually relevant depends on the underlying source. In Table~\ref{table2}, the chiral indices for the different sources are summarized. There are a few interesting things to point out:
\begin{itemize}
\item For most sources, $\bar g_2$ appears at higher order than $\bar g_0$ and/or $\bar g_1$. The one exception is the qEDM, for which $\bar g_{0,1,2}$ are suppressed compared to $\bar d_{0,1}$. Thus, LO calculations of PVTV observables do not require the inclusion of $\bar g_2$.
\item Similar to the PVTC case, the fact that PVTV pion-nucleon interactions without derivatives exist while PVTV $N\!N$ interactions require a derivative, ensures the $N\!N$ potential is dominated by OPE diagrams. The one exception is the gCEDM which conserves chiral symmetry. As a consequence, its contributions to $\bar g_{0,1}$ require a quark mass insertion, such that OPE diagrams and PVTV $N\!N$ interactions appear at the same order \cite{deVries:2011re}.
\item The interactions $\bar d_{0,1}$ describe short-range contributions to the nucleon EDMs. For sources such as $\tb$ and qCEDMs, they carry a chiral index which is larger by two than the corresponding index for $\bar g_0$. As discussed in the next section, this implies that contributions to the nucleon EDMs proportional to $\bar g_0$ (one-loop diagrams) and $\bar d_{0,1}$ (tree level) appear at the same order \cite{Pich:1991fq,Hockings:2005cn,Ottnad:2009jw}. For all other sources, $\bar d_{0,1}$ dominate the nucleon EDMs \cite{deVries:2010ah,Seng:2014pba}. 
\item For most sources the three-pion vertex $\bar \Delta$ appears at higher order than $\bar g_{0}$ and/or $\bar g_1$ and its effects will be minor. For the FQLR, $\bar \Delta$ is relatively enhanced leading to additional contributions to the $N\!N$ potential and a PVTV three-body force \cite{deVries:2012ab}.
\end{itemize}

With the interactions in Eq.~\eqref{Lpvtv} it is straightforward to obtain the PVTV $N\!N$ potential \cite{Maekawa:2011vs}: 
\begin{eqnarray}\label{eq:nnpot}
  V_{\mathrm{PVTV}} = 
   &&i\frac{g_A \bar g_0}{2F_\pi}(\vec{\tau}_{1}\cdot\vec{\tau}_{2})\frac{(\vec \sigma_1 - \vec \sigma_2)\cdot \vec q}{q^2+\mpi^2}+ \frac{i}{2}\left(\bar C_1+\bar C_2\vec{\tau}_{1}\cdot\vec{\tau}_{2}\right)(\vec \sigma_1 - \vec \sigma_2)\cdot \vec q\,.\nn \\
&+&i\frac{g_A}{4 F_\pi}
\left(\bar g_1+\bar \Delta\, \,f_{\bar g_1}(q) \right) \left[(\vec \sigma_1 + \vec \sigma_2)(\tau_1-\tau_2)^3+(\vec \sigma_1 - \vec \sigma_2)(\tau_1+\tau_2)^3 \right]\cdot \frac{\vec q}{\vec{q}^{2}+\mpi^2}\,\,\,,
\end{eqnarray}
where $f_{g_1}(q)$ describes a momentum-dependent correction to the isospin-breaking potential. It arises from a one-loop diagram involving $\bar \Delta$ \cite{deVries:2012ab} and is given by
\begin{equation}\label{eq:g1corr}
f_{\bar g_1}(q) \equiv \, -\frac{15}{32}\frac{g_A^2\mpi m_N}{\pi F_\pi^2}\left[1+ \left(\frac{1+2q^{2}/(4\mpi^2)}{3 q/(2\mpi)}\arctan\left(\frac{q}{2\mpi}\right)-\frac{1}{3}\right)\right]\,.
\end{equation}
Explicit calculations in light nuclei \cite{Bsaisou:2014zwa} show that the dominant part arises from the $q^2$ independent part in which case the $\bar \Delta$ dependence can be absorbed in
$\bar g_1 \rightarrow \bar g_1+\bar \Delta\, \,f_{\bar g_1}(0)$, which we do henceforth. In addition, the $\bar \Delta$ vertex induces a three-body PVTV potential which can influence EDMs of nuclei with $A\geq 3$.  Higher-order contributions to the potential, such as TPE diagrams, have been studied in Refs.~\cite{Maekawa:2011vs,Bsaisou:2012rg}. They are not as interesting as their PVTC counterparts because the $\bar g_0$ TPE diagrams have the same spin-isospin structure as the LO OPE, whereas the sum of $\bar g_1$ TPE diagrams vanishes. 

\subsubsection{Sizes of PVTV LECs and the nucleon EDMs}\label{LECsPVTV}
In order to calculate PVTV observables such as EDMs in terms of the fundamental couplings at the quark level, we must know the sizes of the LECs appearing in Eq.~\eqref{Lpvtv}. The chiral indices in Table~\ref{table2} provide some information about  their relative sizes but this is  based on NDA and not very precise. Various nonperturbative methods have been applied to calculate the LECs, see Ref.~\cite{Engel:2013lsa} for an overview. Nevertheless, for most PVTV sources very little is known which hampers the interpretation of EDM experiments. 

We begin with the PVTV pion-nucleon and $N\!N$ interactions. By far, the most is known for the $\tb$ term. As was realized in Ref.~\cite{Crewther:1979pi}, chiral symmetry relates the PVTV LECs to PCTC LECs originating in the quark masses. For instance, $\bar g_0$ is proportional to $\delta m_N$, the strong part of the proton-neutron mass splitting.\footnote{$SU(3)$ symmetry also relates $\bar g_0$ and mass splittings of octet baryons, but these relations suffer from large $SU(3)$-breaking corrections \cite{deVries:2015una}.}. A similar, but less precise, relation exists between $\bar g_1$ and the strong part of the pion mass splitting \cite{Mereghetti:2010tp,Bsaisou:2014zwa}. With lattice-QCD input for the mass splittings \cite{Walker-Loud:2014iea,Borsanyi:2014jba}, the following values are obtained \cite{Bsaisou:2014zwa,deVries:2015una}:
\begin{equation}\label{thetag0g1}
\bar g_0 = -(15.5\pm 2.5)\cdot 10^{-3}\,\tb,\qquad \bar g_1 = (3.4\pm 2)\cdot 10^{-3}\, \tb\,\,\,,
\end{equation}
such that $|\bar g_1/\bar g_0| \sim 1/5$. Owing to the small value of the proton-neutron mass splitting, this ratio of couplings is somewhat larger than expected from Table \ref{table2} which predicts $|\bar g_1/\bar g_0| \sim \mathcal O(m_\pi^2/\Lambda_\chi^2)$ \cite{Bsaisou:2012rg}. This discrepancy reflects the uncertainty of the NDA estimates in Table \ref{table2}. With similar methods, PVTV couplings between heavier mesons and nucleons can be calculated. For example, the PVTV $\eta$-nucleon coupling is proportional to the nucleon sigma term \cite{deVries:2015una}, which is nowadays determined with high precision \cite{Hoferichter:2015dsa}. Values for $\bar C_{1,2}$ can then be estimated using resonance saturation: $\bar C_1 \simeq (-8 \cdot 10^{-3})\, \tb\, \mathrm{fm^3}$ and $\bar C_2 \simeq (-1 \cdot 10^{-3})\, \tb\, \mathrm{fm^3}$.

Much less is known for the dimension-six operators. In case of the qCEDMs, values of $\bar g_{0,1}$ have been obtained with QCD sum rules \cite{Pospelov:2001ys}, but with $\mathcal O(100\%)$ uncertainties. For the other sources no calculations of $\bar g_{0,1}$ or $\bar C_{1,2}$ exist and at the moment we cannot do better than NDA to estimate their sizes. In some cases, chiral symmetry considerations provide some information about the relative sizes of the couplings. For example, for the FQLR the ratio of couplings is predicted: $\bar g_0/\bar g_1=\delta m_N/(8 c_1 \mpi^2) \simeq 0.01$ \cite{deVries:2012ab}. 

Next we study $\bar d_{0,1}$ which describe short-range contributions to the nucleon EDMs. These terms are renormalized by loop contributions proportional to $\bar g_{0,1}$ and the total EDMs of the neutron, $d_n$, and proton, $d_p$, up to NLO are given by \cite{Crewther:1979pi,Hockings:2005cn,Mereghetti:2010kp} \footnote{The lengthier $SU(3)$ $\chi$PT expressions can be found in Refs.~\cite{Pich:1991fq,Borasoy:2000pq,Ottnad:2009jw}.}
\bea \label{nucleonEDM}
  d_n& = & {\bar d}_0 -\bar d_1-\frac{e g_A\bar g_0}{8\pi^2 F_\pi} \left(  \ln
\frac{m_\pi^2}{\mu^2} -\frac{\pi m_\pi}{2 m_N} \right)\,\,\,,\nn\\
\label{eq:dpfull}
d_p & = & {\bar d}_0 + \bar d_1+\frac{e g_A}{8\pi^2 F_\pi} \left[ \bar g_0 \left(  \ln
\frac{m_\pi^2}{\mu^2} -\frac{2\pi m_\pi}{m_N} \right) -
\bar g_1 \frac{\pi m_\pi}{2 m_N} \right]\,\,\,,
\eea
where $e\!>\!0$ is the proton charge, dimensional regularization ($\overline {\mathrm{MS}}$ scheme) is applied, and $\mu$ denotes the renormalization scale. Which of these terms are relevant depends again on the source, see Table~\ref{table2}. Using $(4\pi F_\pi)\simeq \Lambda_\chi$, we see that for $\tb$ and qCEDMs, both $\bar d_{0,1}$ and $\bar g_0$ contribute at LO. Inserting Eq.~\eqref{thetag0g1} into Eq.~\eqref{nucleonEDM} and $\mu=m_N$, gives
$d_n \simeq \bar d_0 -\bar d_1 - (2\cdot 10^{-16})\,\tb\,e\,\mathrm{cm}$. The current experimental constraint on $d_n$ \cite{Baker:2006ts} then limits $|\tb| \leq 10^{-10}$, assuming no cancellations between $\bar d_0-\bar d_1$ and the loop piece. A more precise constraint requires nonperturbative information about the short-range terms. For all other sources, the nucleon EDMs are dominated by $\bar d_{0,1}$ and chiral symmetry provides little information about their sizes.

Lattice QCD can directly determine the LECs $\bar d_{0,1}$ in terms of the PVTV quark-gluon operators. In recent years, a lot of effort has gone into determining the nucleon EDMs from the $\tb$ term \cite{Shintani:2005xg,Shintani:2008nt}. The simulations take place at nonphysical quark masses and in a finite volume and $\chi$PT expressions are needed to extrapolate to the physical point and infinite volume \cite{O'Connell:2005un,Akan:2014yha}. Based on unpublished lattice data from Shintani et al, Ref.~\cite{Guo:2012vf,Akan:2014yha} extracted the following values
\begin{equation}\label{dndplattice}
d_n =  - (2.7\pm1.2)\cdot 10^{-16}\,e\,\mathrm{cm}\,\, ,\qquad d_p =   (2.1\pm1.2)\cdot 10^{-16}\,e\,\mathrm{cm}\,\,\,.
\end{equation}
Ref.~\cite{Guo:2015tla} performed a lattice-QCD calculation of $d_n$ by analytically continuing $\tb$ into the complex plane. In this way, the Euclidean action becomes real and standard stochastic methods can be applied. The following value was found 
\begin{equation}\label{dndplattice2}
d_n =  - (3.8\pm1.0)\cdot 10^{-16}\,e\,\mathrm{cm}\,.
\end{equation}
Finally, Ref.~\cite{Shindler:2015aqa} performed a quenched calculation of $d_n$ and $d_p$ based on the gradient flow for gauge fields \cite{Luscher:2010iy}, finding results consistent with Eqs.~\eqref{dndplattice} and \eqref{dndplattice2}. $\bar g_0$ was extracted from the calculated electric dipole radius by a comparison to  $\chi$PT predictions \cite{Ottnad:2009jw,Mereghetti:2010kp}, finding a value a few times larger than Eq.~\eqref{thetag0g1}. Considering the limitations of the calculation (quenched and $800$ MeV pion mass), not much can be said about this discrepancy. 

The other well-studied operator is the qEDM. As can be glimpsed from Table \ref{table2}, the only relevant LECs are $\bar d_{0,1}$ as all other interactions are suppressed by $\alpha_{\mathrm{em}}/4\pi$. The nucleon EDMs can be related to the nucleon tensor charges and a recent lattice QCD calculation \cite{Bhattacharya:2015esa} found
\begin{equation}\label{dndq}
d_n = -(0.23\pm0.03)d_u +(0.77\pm0.07)d_d\,,\qquad d_p = (0.77\pm0.07)d_u -(0.23\pm0.03)d_d\,\,\,.
\end{equation}

Much less is known for the remaining dimension-six sources. For the qCEDMs, the nucleon EDMs have been calculated with QCD sum rules with $50\%$ accuracy \cite{Pospelov:2000bw,Hisano:2012sc}, while for the gCEDM and four-quark operators only estimates at the order-of-magnitude level exist \cite{Weinberg:1989dx,Demir:2002gg,deVries:2010ah,Seng:2014pba}. Improvements on the hadronic matrix elements could have significant impact on falsifying or identifying specific BSM scenarios from EDM measurements, see Refs.~\cite{Inoue:2014nva,Dekens:2014jka, Chien:2015xha} for discussions. 

\section{Discrete symmetry breaking in few-body systems}\label{observables}
Having discussed the discrete-symmetry-breaking $N\!N$ potentials, we now turn to observables in few-body systems. We mainly focus on systems with two nucleons as these capture the essential ideas and avoid the technical difficulties associated with larger systems.  
The starting point of the calculations is the PCTC $N\!N$ strong interaction from which nuclear wave functions are calculated. 
In most of the literature on discrete symmetry breaking, phenomenological high-quality potentials such as Argonne A$v_{18}$ \cite{Wiringa:1994wb} or NijmegenII \cite{Stoks:1994wp} are used which accurately describe the $N\!N$ experimental data-base. The obtained wave-functions are then combined with PVTC/PVTV  OBE models (such as the DDH model) or with the $\chi$EFT potentials. The latter method is usually called the hybrid approach.

Calculations where also the wave functions are obtained from $\chi$EFT have been performed in various frameworks. At very low energies, the pion can be integrated out. In the resulting pionless EFT \cite{Bedaque:2002mn}, all interactions are described by $N\!N$ contact vertices which allows for much simpler calculations. The absence of pions implies that the EFT only converges in the low-energy region $E\sim M_\pi^2/(2 m_N )\simeq 10$\,MeV. By integrating out the pion, the hierarchy of forces due to the chiral-symmetry properties of the PVTC and PVTV quark-gluon operators is obscured. The pionless approach has been applied to several PVTC observables \cite{Phillips:2008hn,Griesshammer:2010nd,Schindler:2009wd,Griesshammer:2011md,Vanasse:2014sva} and is reviewed in Ref.~\cite{Schindler:2013yua}.

Another EFT approach is based on the KSW (after the authors of Ref.~\cite{Kaplan:1998xi}) power counting. Within this framework, the LO PCTC $N\!N$ contact interactions are resummed to obtain scattering lengths and bound-state properties. Pions are treated in perturbation theory which has the advantage that analytic calculations become possible and renormalization is explicit. Several PVTC and PVTV observables in the two-body sector have been calculated in this way \cite{Savage:1998rx,Kaplan:1998tg,deVries:2011re}. Unfortunately, higher-order calculations \cite{Fleming:1999ee} show that the KSW expansion already fails to converge at fairly low momenta $q\sim m_\pi$ such that observables associated with larger momentum transfer or denser nuclei cannot be treated in this fashion.

Here we mostly discuss the approach, which we will call $\chi$EFT, where pions are treated explicitly and nonperturbatively and both PCTC and symmetry-breaking potentials are obtained from $\chi$EFT.  Typically numerical differences between hybrid and $\chi$EFT calculations are small if observables are mainly determined by long-range physics (i.e. pions). Although the PCTC $\chi$EFT potential can be calculated in perturbation theory, the potential itself must be iterated to all orders to calculate $N\!N$ scattering lengths, phase shifts, and bound state properties. 
Technically, the task is to solve the non-relativistic Lippmann-Schwinger (LS) equation
\begin{eqnarray}\label{LS}
T = V + VG_0 T\,\,\,,
\end{eqnarray}
written here in short-hand notation. $T$ denotes the scattering matrix, $G_0$ the non-relativistic propagator of the free nucleon pair, and $V$ the PCTC strong potential. Once $T$ is obtained numerically, the effects of the PVTC and PVTV potentials can be calculated in perturbation theory. 
 In practice, it can be easier to solve Eq.~\eqref{LS} directly with $V$ the sum of the strong and symmetry-breaking potentials. Because the latter are proportional to very small parameters, their iteration leads to numerically identical results as first-order perturbation theory. Details on the solution of the LS equation in the presence of $P$ violation can be found in Refs.~\cite{Carlson:2001ma,deVries:2013fxa}.
The main consequence of the PVTC and PVTV potentials is that transitions between states with even and odd orbital angular momentum become possible. For instance, the PVTC OPE potential in Eq.~\eqref{onepion} leads to ${}^3S_1 \leftrightarrow {}^3P_1$ transitions normally forbidden by P conservation. 

In general, the momentum integral in the LS equation is divergent and needs to be regulated. Different choices are of course possible and most results below are based on Ref.~\cite{Epelbaum:2004fk} which regularized 
the LS equation by multiplying the total potential by the function\footnote{Ref.~\cite{Entem:2003ft} applies a similar function but uses $n=2$ for NLO and higher-order corrections. The PVTC $\chi$EFT analysis of Ref.~\cite{Viviani:2014zha} uses a different regulator for the PVTC potential.}
\begin{equation}
V(p^\prime, p) \to
\mathrm{exp}[-p^{\prime\,2n}/\Lambda^{2n}]\,\, V
(p^\prime, p)
\,\,\mathrm{exp}[-p^{2n}/\Lambda^{2n}]~, 
\end{equation}
where $n=3$ and $\Lambda$ is a momentum cut-off. Recently, a regularization scheme in coordinate space has been developed which better preserves the long-range part of the pion-exchange terms \cite{Epelbaum:2014efa}. This scheme has not been applied yet to PVTC or PVTV calculations. Much has been written about the size of the momentum cut-off, $\Lambda$, \cite{Nogga:2005hy,Phillips:2013fia}, which we cannot discuss in detail here. Although 
$\Lambda$ can in principle be any high-energy scale, for practical purposes it is optimal  to pick $\Lambda$ of similar size as $\Lambda_\chi$ \cite{Entem:2003ft,Epelbaum:2004fk}. The $\chi$EFT calculations discussed below all varied $\Lambda$ between $450$ and $600$ MeV to get an estimate of (part of the) theoretical uncertainties. 
 
 \subsection{Few-body PVTC processes}
 We now turn to the calculations of PVTC observables. For an overview of PVTC experiments we refer to Ref. \cite{Haxton:2013aca}. We focus on the observables in the two-body system\footnote{We do not discuss observables in the one-body sector such as the proton anapole moment \cite{Maekawa:2000qz} that vanished for on-shell photons but provides a radiative correction to PVTC electron-proton scattering \cite{Musolf:1990sa}.} for which nonzero signals have been measured. Afterwards, we briefly discuss more complicated systems. 
 
 \subsubsection{Proton-proton scattering}
The first observable we discuss is the $pp$ longitudinal analyzing power (LAP)  which vanishes if $P$ is conserved. It is defined as the difference in cross section of scattering between an unpolarized target and a beam of positive and negative helicity, normalized to the sum of cross sections. The existing experiments measured the LAP over a certain angular range (from $\theta_1$ to $\theta_2$) and report the integrated asymmetry:
\begin{equation}
\bar A_L(E,\theta_1,\theta_2) = \frac{\int_{\theta_1}^{\theta_2} d\Omega(d\sigma_+ - d\sigma_-)}{\int_{\theta_1}^{\theta_2} d\Omega(d\sigma_+ + d\sigma_-)}\,\,\,.
\end{equation}
$A_L$ has been measured for beam energies of $13.6$~MeV \cite{Eversheim:1991tg},  $45$~MeV 
\cite{Kistryn:1987tq}, and $221$ MeV\cite{Berdoz:2001nu}. The experiments at $13.6$ and $45$ MeV are scattering experiments which measured the $pp$ LAP over an angular 
range of, respectively, $20^{\circ}$-$78^{\circ}$ and $23^{\circ}$-$52^{\circ}$ (lab coordinates)
\begin{eqnarray}\label{exp12}
\bar A_L (13.6\,\mathrm{MeV}) = (-0.93\pm 0.21)\cdot 10^{-7}~,\qquad\bar A_L (45\,\mathrm{MeV}) = (-1.50\pm 0.22)\cdot  10^{-7}~.
\end{eqnarray}
 The experiment at $221$~MeV is a transmission experiment and effectively measures the LAP over the whole angular range modulo a small opening angle \cite{Driscoll:1988hg}
\begin{eqnarray}\label{exp3}
\bar A_L (221\,\mathrm{MeV}) &=& (0.84\pm0.34)\cdot 10^{-7}~.
\end{eqnarray}

Traditionally, it was taken that  the $pp$ LAP does not depend on the weak pion-nucleon coupling $h_\pi$ \cite{Driscoll:1988hg, Carlson:2001ma} as the OPE potential does not contribute to $pp$ scattering. In $\chi$EFT, the first non-vanishing contributions appear at NLO and consist of TPE contributions and short-range $N\!N$ interactions. The former depend on $h_\pi$ and the latter on the combination of LECs $C= -C_0+C_1+C_2-C_3$ that gives rise to ${}^1S_0 \leftrightarrow {}^3 P_0$ transitions. 

The calculation of $\bar A_L$ in terms of $h_\pi$ and $C$ is complicated by the Coulomb interaction. This problem has been carefully studied in Refs.~\cite{Carlson:2001ma,deVries:2013fxa}. The low-energy scattering experiments do not measure over small angles and the Coulomb interactions plays a minor role. The transmission experiment is more affected. The energy of $221$ MeV was chosen because the contribution from $j=0$ transitions are proportional to the sum of the strong phase shifts $\delta_{{}^1 S_0}+\delta_{{}^3 P_0}$ which vanishes around $220$ MeV \cite{Driscoll:1988hg,Driscoll:1989jv}. Therefore $\bar A_L(221\,\mathrm{MeV})$ is sensitive to higher partial waves (mostly ${}^3P_2 \leftrightarrow {}^1D_2$ transitions that are proportional to $h_\pi$) and is complementary to the low-energy experiments. However, the Coulomb interactions shift the vanishing of the $j=0$ phase shifts to a somewhat larger energy \cite{deVries:2013fxa}. Combined with the larger uncertainty of the $\chi$EFT potentials at higher energies, the data at $221$ MeV unfortunately does not pin down a precise value for $h_\pi$.

The LECs were fitted\footnote{The analogous analysis of DDH parameters was performed in Ref.~\cite{Carlson:2001ma}.} in Refs.~\cite{deVries:2013fxa} and \cite{Viviani:2014zha} using, respectively, N${}^3$LO strong potentials from Refs.~\cite{Epelbaum:2004fk} and \cite{Entem:2003ft}. Ref.~\cite{deVries:2013fxa} found the strongly correlated (see Fig.~\ref{FigPV}) values
 \begin{eqnarray}\label{fit3}
h_\pi = (1.1\pm 2 )\cdot 10^{-6}~,\qquad
C = (-6.5\pm 8)\cdot 10^{-6}~,
\end{eqnarray}
in excellent agreement with Ref.~\cite{Viviani:2014zha}. The uncertainty is almost completely determined by the lack of data and the experimental uncertainties, while uncertainties due to cut-off variations are much smaller. An additional experiment around $125$ MeV would be very beneficial in reducing the uncertainties of the fit \cite{deVries:2013fxa}. 
Ref.~\cite{deVries:2014vqa} investigated the contributions to $\bar A_z$ from the N${}^2$LO PVTC potential. As discussed in Sec.~\ref{PVTCpot}, unless $h_\pi$ is very small, the dominant part of the N${}^2$LO potential is expected to be proportional to $c_4 h_\pi$. Including this correction, the fit slightly improves
$h_\pi = (0.8\pm 1.5 )\cdot 10^{-6}$ and $C = (-5.5\pm 7)\cdot 10^{-6}$, but does not strongly affect the values of the LECs indicating that the $\chi$EFT expansion is converging satisfactorily. The other parts of the N${}^2$LO potential depend on additional PVTC LECs and more data is needed before they can be fitted.

 \subsubsection{Neutron capture on the proton}
We now discuss the LAP, $a_\gamma$, for the process\footnote{The inverse process $\vec \gamma d \rightarrow np$ could provide complementary information, see for instance Refs.~\cite{Schiavilla:2004wn,Vanasse:2014sva}.} $\vec n p\rightarrow d \gamma$ which is defined as 
\bea\label{LAP}
A_\gamma(\theta) & = &\frac{d\sigma_+(\theta) - d \sigma_-(\theta)}{d\sigma_+(\theta) + d \sigma_-(\theta)} = a_\gamma\,\cos \theta\,\,\,,
\eea
with $d\sigma_\pm(\theta)$ the differential cross section for neutrons with positive/negative helicity and $\theta$ the angle between the photon momentum and the neutron spin. For a long time only upper bounds existed for $a_\gamma$ \cite{Cavaignac:1977uk,Gericke:2011zz}, but recently a  preliminary result was reported~\cite{Crawford} 
\begin{equation}\label{agammaEXP}
a_\gamma = (-7.14\pm4.4)\cdot 10^{-8}\,\,\,,
\end{equation}
only two standard deviations away from zero. 
The result is based on a subset of the data taken by the NPDGamma collaboration. A full result with uncertainty $\sim10^{-8}$ is expected soon.

The great interest in measuring $a_\gamma$ is that, in contrast to $\bar A_z$, it does depend on the LO PVTC potential and is therefore more sensitive to $h_\pi$. It has been studied in detail in the DDH framework \cite{Schiavilla:2002uc}, hybrid calculations \cite{Hyun:2001yg,Liu:2006dm}, pionless EFT \cite{Schindler:2009wd}, KSW-counting \cite{Kaplan:1998xi}, and recently $\chi$EFT \cite{deVries:2015pza}. A new aspect of $a_\gamma$ is that it depends on PCTC and PVTC electromagnetic currents, which can be calculated along the same lines as the $N\!N$ potentials. The experiment takes place at thermal energies where the capture cross section is dominated by the nucleon isovector magnetic moment, while NLO two-body currents contribute at the $5\%$ level. In combination, with the N${}^3$LO $\chi$EFT potential\footnote{Note the inconsistenty in the chiral order of the potential and currents as is typical in these kind of calculations. Consequently, the uncertainty is dominated by missing higher-order currents.} \cite{Epelbaum:2004fk}, these currents predict a capture cross section \mbox{$319\pm 5$ mb} \cite{deVries:2015pza}, compared to the experimental \mbox{$334.2\pm0.5$ mb} \cite{Cox1965497}. The remaining $4\%$ discrepancy should be due to higher-order effects such as TPE currents \cite{Kolling:2009iq}. These have not been included in the analysis as the corresponding PVTC TPE currents have not been calculated.   

Apart from the above ingredients, the LO calculation of $a_\gamma$ requires the nucleon convection current (from gauging the nucleon kinetic energy) and PVTC OPE currents  proportional to $h_\pi$. $a_\gamma$ then arises from an interference between the isovector magnetic moment and:
1) the convection current and the LO PVTC potential, 2) the PCTC OPE currents and the  LO PVTC potential, 3) the PVTC OPE currents.
The individual contributions only have a minor dependence on the cut-off used to regulate the LS equation, but the sum suffers from a relatively larger dependence due to cancellations \cite{Hyun:2001yg,deVries:2015pza}. Ref.~\cite{deVries:2015pza} found
\begin{equation}
a_\gamma = -(0.11\pm0.05)\,h_\pi\,\,\,.
\end{equation}
The central value agrees with DDH and hybrid results \cite{Schiavilla:2002uc, Liu:2006dm} that applied the Siegert theorem to relate the electric dipole currents to the one-body charge destiny. The observed cutoff dependence is likely to be reduced by the inclusion of higher-order corrections to the PCTC and PVTC exchange currents\footnote{If $h_\pi$ turns out to be very small,  the OPE and TPE currents are suppressed and the dominant contribution $a_\gamma$ would depend only the PVTC contact LEC $C_4$: $a_\gamma = -(0.055\pm0.025)\,C_4$ \cite{deVries:2015pza}. A contribution at the same order from a PVTC pion-nucleon-photon interaction is negligible \cite{Liu:2006dm}. }. Although these can be partially taken into account by the Siegert theorem, this approach most likely overestimates the accuracy of the calculation as not all exchange currents are taken into account. 

 \begin{figure}[!t]
\centering
\includegraphics[width=0.45\textwidth]{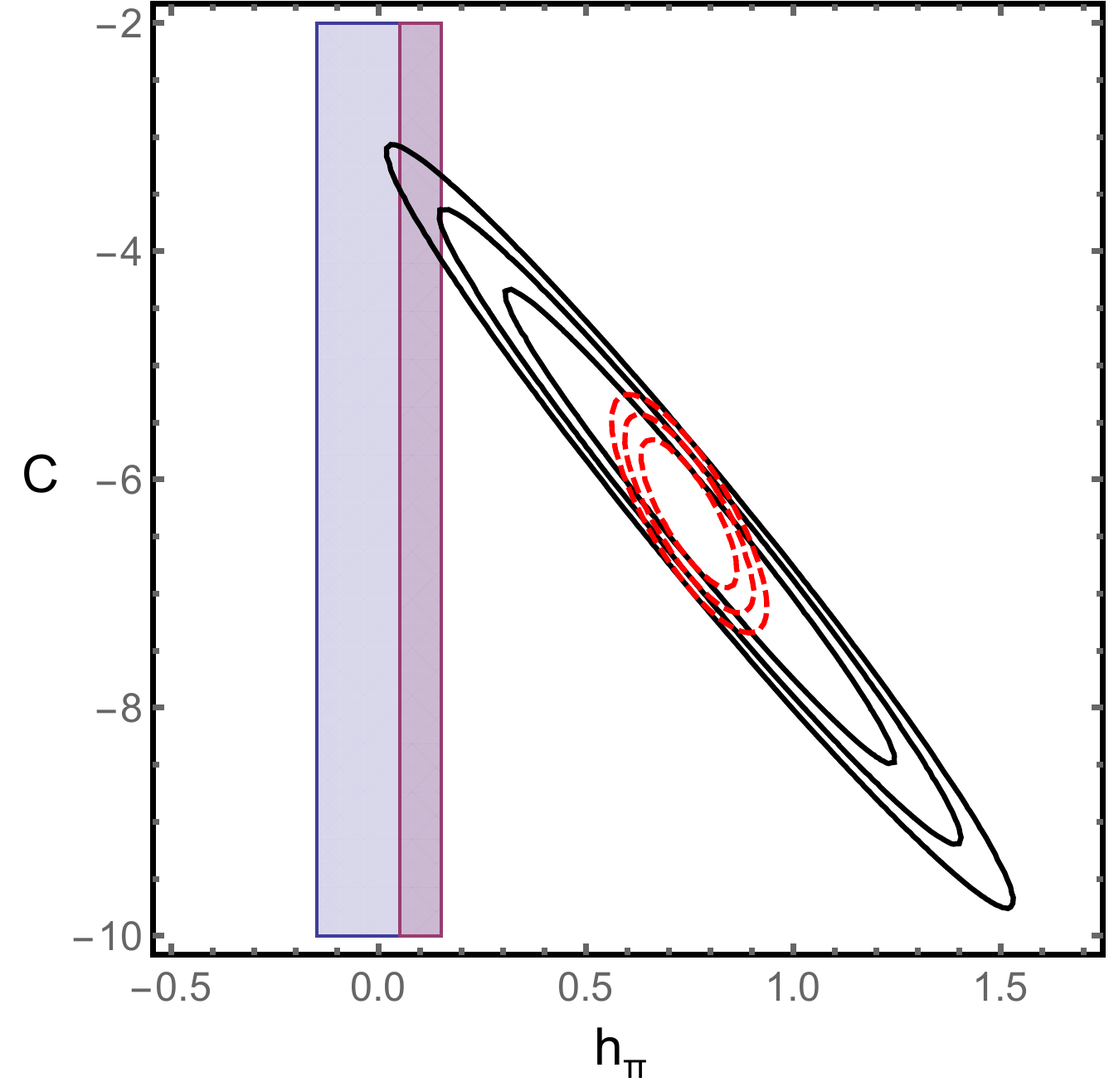} 
\caption{\small The allowed ranges for the LECs $h_\pi$ and $C$ (in units 
of $10^{-6}$). The black ellipses correspond to fits to the combined $pp$ and $np$ data with total $\chi^2=\{2,3,4\}$. The red (dashed) ellipses are the same, but using the expected future experimental uncertainty (at $10^{-8}$ level) for the $\vec n p\rightarrow d \gamma$ measurement, while keeping the central value of Eq.~\eqref{agammaEXP}. The shaded regions shows the constraints on $h_\pi$ from ${}^{18}$F data and a lattice-QCD calculation of $h_\pi$. 
}
 \label{FigPV}
\end{figure}

The analysis of $\bar A_L$ and $a_\gamma$ can now be combined to extract more precise values of $h_\pi$ and $C$. Contours of total $\chi^2=2,3,4$ ($\chi_\mathrm{min}\simeq 0.7)$ in the $h_\pi-C$ plane are plotted\footnote{The plot is based on based on the N${}^3$LO $\chi$EFT potential \cite{Epelbaum:2004fk} with intermediate cut-off values: $\Lambda = 550$ MeV  and $\Lambda_S = 600$ MeV.}  in Fig.~\ref{FigPV}. Varying the cut-offs and considering the whole allowed range, Ref.~\cite{deVries:2015pza} conservatively extracted
\begin{equation}
h_\pi = (1.1 \pm 1.0)\cdot 10^{-6}\,\,\,,\qquad C = (-6.5\pm 4.5)\cdot 10^{-6}\,\,\,.
\end{equation}
The fits indicate that small values of $h_\pi \sim 10^{-7}$ are still consistent with the data, but that larges values of $h_\pi\sim(5-10)\cdot 10^{-7}$ are preferred.  Such values are in conflict with the bound on ${}^{18}$F gamma-ray emission, $h_\pi \leq 1.3\cdot 10^{-7}$, and lattice and model calculations of $h_\pi \simeq 10^{-7}$. The upcoming increase in sensitivity of the $a_\gamma$ measurement will tell whether small values of $h_\pi$ are consistent with few-body experiments as can be seen from the red (dotted) ellipses. 

\subsubsection{Additional PVTC observables}
A fairly large set of few-body PVTC observables could provide additional input on nuclear PVTC \cite{Schindler:2013yua}. Examples in two- and three-body cases are the spin rotation in $\vec n p$ or $\vec n d$ and the LAPs of $\vec n d$ scattering and $\vec n d \rightarrow {}^3\mathrm{H} \gamma$, which have been discussed in DDH \cite{Schiavilla:2004wn,Song:2012yx}, pionless EFT \cite{Griesshammer:2011md}, hybrid calculations \cite{Liu:2006dm}, and, recently in $\chi$EFT \cite{Viviani:2014zha}.  The $\chi$EFT analysis\footnote{Ref.~\cite{Viviani:2014zha} uses a different parametrization of the PVTC $N\!N$ LECs. The results in Eqs.~\eqref{spinrot} and \eqref{chargeexchange} have been translated to Eq.~\eqref{contactPV}  via the relations: $C_1^{\mathrm{Viv}}=C_0/2-C_1$, $C_2^{\mathrm{Viv}}=C_0/2$, $C_3^{\mathrm{Viv}}=-C_4$, $C_4^{\mathrm{Viv}}=-C_2/2$, and $C_5^{\mathrm{Viv}}= C_3/2$, where $C_i^{\mathrm{Viv}}$ denotes the LECs appearing in Ref.~\cite{Viviani:2014zha}.} \cite{Viviani:2014zha} of the spin rotation angles gives
\begin{eqnarray}\label{spinrot}
\frac{d\phi}{dz}(\vec np)&=& (1.31\pm0.05)h_\pi + (0.20\pm 0.01) C_0 -(0.23\pm0.01) C_1 - (0.44\pm0.01)C_3 - (0.09\pm0.01)C_4\nonumber\\
\frac{d\phi}{dz}(\vec nd)&=& (2.20\pm0.02)h_\pi - (0.08\pm 0.01) C_0 - (0.19\pm0.01)C_4\,\,\,,
\end{eqnarray}
in units of $\mathrm{rad}/m$. 
Another interesting observable is the LAP of the reaction $\vec n + {}^3\mathrm{He}\rightarrow  p +{}^3\mathrm{H}$:
\begin{equation}\label{chargeexchange}
A_L = -(0.14\pm0.01)h_\pi + (0.017\pm 0.003) C_0 -(0.007\pm 0.001) C_1+ (0.008\pm0.001) C_2 + (0.018\pm0.002)C_4\,\,\,.
\end{equation}
These results  \cite{Viviani:2014zha} are based on the PCTC potential of Ref.~\cite{Entem:2003ft} supplemented by N${}^2$LO three-body forces \cite{Epelbaum:2002vt}. The $h_\pi$ dependence is dominated by the OPE potential, with TPE contributions entering at the $10\%-30\%$ level as expected from power counting.  The spin rotation angles have not been measured yet, but experiments are being considered. The LAP $A_L$ is planned to be measured with sensitivity $\sim1.6\cdot 10^{-8}$ \cite{Crawford2}. 

Finally, we mention the LAP in $\vec p \alpha$ scattering for which a nonzero value has been measured $A_L(46\,\mathrm{MeV}) = -(3.3\pm0.9)\cdot 10^{-7}$ \cite{Lang:1985jv}. No $\chi$EFT calculation exists for this process, which is unfortunate as the measurement could have important implications for the extraction of the PVTC LECs. Ref.~\cite{Roser:1985rs} found $A_L = -0.34 h_\pi + \dots$, based on an optical model and the dots denote contributions from other DDH parameters. The difficulty in the $\chi$EFT calculation lies in the five-body problem and the presence of Coulomb interactions. The great progress in ab initio few-body calculations in recent years, see for instance the calculation of $\alpha \alpha$ scattering with nuclear lattice EFT \cite{Elhatisari:2015iga}, might make it possible to analyze the $\vec p \alpha$ LAP with similar methods. The same applies for the $\vec n \alpha$ spin rotation \cite{Snow:2011zza}.

 \subsection{PVTV effects in few-nucleon systems.}
The main motivation for the study of PVTV in few-nucleon systems are the plans to measure EDMs of  light nuclei in storage rings. 
Traditional EDM experiments essentially consist in looking for a change in the spin precession of the system in the presence of electromagnetic fields. A charged particle at rest in an electric field would quickly escape the experimental apparatus. This is not true for charged particles confined in a  storage ring and it was realized that EDMs of charged particles could be measured in such a setup \cite{Farley:2003wt}.
The JEDI collaboration \cite{JEDI,Eversmann:2015jnk} plans to measure EDMs of light ions\footnote{Schiff's theorem implies that the nuclear EDM
in an atomic system is screened by the surrounding electrons. The screening is very effective for light atoms, hence the need to measure the EDMs of light nuclei directly. } such as the proton, deuteron, and helion (${}^3\mathrm{He}$ nucleus) in a dedicated storage ring with an accuracy exceeding current neutron EDM experiments. 

The advantage of going to systems with more than one nucleon, is that their EDMs become sensitive to the PVTV potential in Eq.~\eqref{eq:nnpot}. The nucleon EDMs depend on the PVTV pion-nucleon LECs only via loop diagrams associated with the typical $\mpi^2/(4\pi F_\pi)^2$ suppression factor.  The EDMs of nuclei, however, depend already at tree level on the PVTV pion-nucleon LECs such that the EDMs of the bound states can be considerably larger than the constituent nucleon EDMs \cite{Sushkov_Flambaum_Khriplovich_1984,Lebedev:2004va,deVries:2011re}. As discussed below, this argument only holds for some of the PVTV sources at the quark-gluon level such that combined measurements of EDMs of nucleons and light nuclei would point towards the underlying PVTV source \cite{deVries:2011re,deVries:2011an}. The second advantage of light ions is that their EDMs can be calculated to high precision. Calculations of heavier systems such as ${}^{199}$Hg and ${}^{225}$Ra rely on nuclear models and suffer from larger uncertainties. These interesting observables are outside the reach of current $\chi$EFT methods and we refer to Ref.~\cite{Engel:2013lsa} for a detailed discussion.

PVTV effects in processes such as $np$ \cite{Liu:2006qp} and $nd$ \cite{Song:2011sw} scattering can be studied along analogous lines.
However, these effects are most likely too small to be measurable. For instance, Ref.~\cite{Liu:2006qp} concluded that with state-of-the art experimental accuracies, measuring the PVTV  $np$ spin rotation would fall short of the current neutron EDM sensitivities by roughly three orders of magnitude. 

\subsubsection{The EDM of the deuteron and helion}
The prospect of measuring EDMs of light nuclei in storage rings, has led to a number of investigations of the deuteron EDM \cite{Khriplovich:1999qr, Liu:2004tq, deVries:2011re, deVries:2011an,Bsaisou:2012rg}. From the theory point of view, the deuteron is interesting because its spin-isospin properties ensure that its EDM has rather distinctive properties. Most calculations are based on the three PVTV pion-nucleon LECs in Eq.~\eqref{g012} and phenomenological models of the PCTC $N\!N$ potential. A calculations in $\chi$EFT using N${}^2$LO PCTC potentials obtained \cite{Bsaisou:2014zwa}
\begin{equation}\label{dD}
d_D = (0.94\pm0.01)(d_n + d_p) + (0.18\pm0.02)\, \bar g_1\,e\,\mathrm{fm}\,\,\,,
\end{equation}
in good agreement with model and hybrid calculations \cite{Liu:2004tq,deVries:2011an} and a calculation based on KSW counting \cite{deVries:2011re}. The independence on $\bar g_0$ and $\bar C_{1,2}$ can be understood from the fact that the deuteron is mostly in a ${}^3S_1$ state. After an insertion of the terms in Eq.~\eqref{eq:nnpot} proportional to $\bar g_0$ or $\bar C_{1,2}$, the wave function obtains a small ${}^1P_1$ component. 
 The dominant current arises from a coupling to the proton charge and is spin-independent. It therefore cannot return the wave function to its 
${}^3 S_1$ ground state and the contributions vanish. A recent calculation of the ${}^6$Li EDM based on a cluster model found a similar effect and a twice larger dependence on $\bar g_1$ \cite{Yamanaka:2015qfa}. Refs.~\cite{deVries:2011an, Bsaisou:2012rg} investigated higher-order contributions to $d_D$ proportional to $\bar g_0$  from  isospin breaking and higher-order currents, but these were found to be negligible. 

The EDMs of ${}^3$He and ${}^3$H have been calculated in Refs.~\cite{Stetcu:2008vt, deVries:2011an,Song:2012yh,Bsaisou:2014zwa}.  Ref.~\cite{Bsaisou:2014zwa} used the same $\chi$EFT potentials as used for $d_D$ (supplemented by chiral three-body forces) and found\footnote{We do not give the triton EDM results as it most likely will not be measured due to its radioactive nature.}:
\begin{eqnarray}\label{d3He}
d_{{}^3{\rm He}} &=& (0.90 \pm 0.01)\, d_n-(0.03\pm 0.01 )\,d_p\nn \\
&&\left[- (0.017\pm0.006)\,\bar \Delta + (0.11\pm 0.01)\,\bar g_0 + (0.14\pm 0.02)\,\bar g_1 \right]e\, {\mathrm{fm}} \nn\\
&& \left[ -(0.04\pm 0.02)\bar C_1 + (0.09\pm 0.02)\bar C_2 \right] \,e\,{\mathrm{fm}}^{-2} \, \,\, . \label{eq:he3edm}
\end{eqnarray}
Apart from a smaller contribution from $d_p$ and a tiny dependence on $\bar \Delta$ via the PVTV three-body force\footnote{Power counting \cite{deVries:2012ab,Bsaisou:2014zwa} predicted a larger dependence on $\bar \Delta$ and it is unclear why the explicit calculation gives a small value.}, the most important difference with respect to $d_D$ is 
 that there is no isospin selection such that $\bar g_0$ and $\bar C_{1,2}$ contribute. It is this observation that makes a $d_{{}^3{\rm He}}$ measurement complementary to that of $d_D$.  The dependence on $\bar C_{1,2}$ in Eq.~\eqref{d3He} is larger than found in Ref.~\cite{deVries:2011an}, possibly due to the pronounced short-range repulsion of the A$v_{18}$ potential used in that work. The uncertainties given in Eqs.~\eqref{dD} and \eqref{d3He} are associated to the nuclear theory and are significantly smaller than uncertainties associated to the hadronic theory, \textit{i.e.}, the sizes of the LECs in terms of the PVTV quark-gluon operators. This is a big advantage over heavier systems where the nuclear uncertainty can be the limiting factor. 

The size of $d_D$ and $d_{{}^3{\rm He}}$ with respect to $d_n$ and $d_p$ depends on the PVTV source. Because the nucleon EDMs induced by the $\tb$ term are mostly isovector, see Eq.~\eqref{dndplattice}, the sum of $0.94(d_n + d_p)=-(0.6\pm1.6)\cdot 10^{-16} \,\tb \,e\,\mathrm{cm}$ appearing in $d_D$ is uncertain. The two-body contribution is of comparable size and could potentially cancel against the one-body terms. We therefore focus on the two-body contributions and by inserting the values of the LECs in Sect.~\ref{LECsPVTV}, we obtain \cite{Bsaisou:2014zwa}  
\begin{eqnarray}
d_D - (0.94\pm0.01)(d_n + d_p) &=&\phantom{-}(0.9\pm0.3)\cdot 10^{-16} \,\tb \,e\,\mathrm{cm}~,\\
d_{{}^3{\rm He}} - (0.90 \pm 0.01)\, d_n + (0.03\pm 0.01 )\,d_p &=& -(1.0\pm0.4)\cdot 10^{-16} \,\tb \,e\,\mathrm{cm}~,
\end{eqnarray}
which provides a clean way to extract $\tb$ from measurements of $d_n$, $d_p$, $d_D$, or $d_{{}^3{\rm He}}$. The short-range operators $\bar C_{1,2}$ contribute at the $10\%$ level as expected from power counting. With lattice calculations of $d_n$ and $d_p$, these relation can be used to test for the existence of the $\tb$ term given measurements of $d_n$ or $d_p$ in combination with $d_D$ or $d_{{}^3{\rm He}}$. 

For the qCEDM, $d_D$ and $d_{{}^3{\rm He}}$ are expected to be dominated by the $\bar g_{0,1}$ terms. Due to the lack of knowledge about the sizes of the LECs, the exact enhancement of $d_D/(d_n+d_p)$ is unclear. QCD sum rules and NDA predict $d_n + d_p$ ($d_n$) to be roughly $10\%$ ($30\%$) of the pion-exchange contribution to $d_D$ ($d_{{}^3{\rm He}}$), but more precise 
statements require lattice input of the PVTV LECs. 

The story is similar for the FQLR but now $\bar g_0$ is suppressed. In this case, the constituent nucleon EDMs are expected to enter at the $10\%$ level such that the ratio $d_{{}^3{\rm He}}/d_D\simeq 0.8$ can be predicted. Such a signal would be a tell-tale sign of existence of the FQLR which is induced in left-right symmetric extensions of the SM (see Ref.~\cite{Dekens:2014ina} and references therein). 

In case of the qEDM, the situation is simple and the light-nuclear EDMs are dominated by the nucleon EDMs. That is, $d_D \simeq d_n+d_p$ and $d_{{}^3{\rm He}} \simeq d_n$. 

Finally, in case of the gCEDM the situation is most complicated. Power counting indicates that all contributions to $d_D$ and $d_{{}^3{\rm He}}$  appear at the same order (apart from $\bar \Delta$). Explicit calculations find a somewhat smaller dependence on $\bar g_{0,1}$ and $\bar C_{1,2}$ then expected \cite{deVries:2011an,Bsaisou:2014zwa}, but nevertheless it is hard to make predictions. For instance, even the relative sign of $d_n$ and $d_p$ is unknown which strongly impacts the interpretation of $d_D$. Lattice input is direly needed.

The above discussion shows that measurements of EDMs of light nuclei could isolate the underlying PVTV source. A different question is whether this information can be used to learn something about the possible SM extension at high energies. In Ref.~\cite{Dekens:2014jka}, various popular scenarios of BSM physics, such as two-Higgs-doublet, left-right symmetric, and supersymmetric models were investigated in the context of EDMs. It was argued that measurements of the EDMs of a few systems could not only distinguish such models from a SM $\tb$ term, but also partially separate the models based on the EDM hierarchy they induce. The analysis could be significantly improved with lattice calculations of the PVTV LECs. 

\section{Final remarks}\label{outlook}
We have reviewed the breaking of discrete space-time symmetries in strongly-interacting systems. We have focused on flavor-diagonal PVTC and PVTV interactions which have very different origin and experimental signatures. The SM weak interaction induces flavor-diagonal PVTC four-quark interactions whose structure are well understood. Flavor-diagonal PVTV interactions appear only in the SM in the form of the QCD $\tb$ term which is strongly constrained by neutron EDM measurements. This smallness leaves room for PVTV effects from BSM physics which at low energies can be parametrized by various interactions of dimension six.  Although in both cases the form of the discrete-symmetry-breaking operators at the quark-gluon level is clear, their manifestation at low energies is obscured by nonperturbative QCD.

To overcome the problem of low-energy QCD, we have focused on the application of $\chi$EFT which allows for the systematic derivation of interaction among the relevant low-energy degrees of freedom: pions, nucleons, and photons. The universality of this approach is reflected by the fact that both PVTC and PVTV chiral Lagrangians and $N\!N$ potentials can be constructed by essentially equivalent methods. The symmetry-breaking potentials can nowadays be combined with PCTC $\chi$EFT potentials to consistently calculate PVTC and PVTV observables in few-nucleon systems. 

In the PVTC case, the LO potential consists of a single interaction proportional to the weak pion-nucleon coupling, $h_\pi$, whose size has been an outstanding issue for a long time. The existing data on PVTC in $\vec p p$ scattering and $\vec n p\rightarrow d \gamma$ allow for an extraction
$h_\pi = (1.1\pm1.0)\cdot 10^{-6}$. This is unfortunately not precise enough to rule out or identify the small values of $h_\pi \leq 10^{-7}$ that are calculated in lattice QCD and preferred by experiments on ${}^{18}$F. The upcoming data on $\vec n p\rightarrow d \gamma$ will hopefully resolve this issue. In any case, additional measurements combined with consistent calculations are needed to confirm the size of $h_\pi$ and fix the values of the LECs appearing in the PVTC potential. 
The goal is to finally get a consistent picture of PVTC nuclear forces.

In the PVTV case, the hierarchy of the $N\!N$ potential crucially depends on the underlying PVTV source. For sources that break chiral symmetry, such as the $\tb$ term or BSM sources like quark chromo-EDMs and the FQLR, the potential is dominated by OPE similar to the PVTC case. The sources can be differentiated by the relative sizes of the PVTV pion-nucleon LECs. The three chiral-breaking sources predict, respectively, $|\bar g_0/\bar g_1| \simeq 5$, $|\bar g_0/\bar g_1| \simeq 1$, $|\bar g_0/\bar g_1| \simeq 0.01$. We have discussed how measurements of the EDMs of light nuclei can be used to probe these ratios and disentangle the sources. For other sources, such as the gCEDM, the PVTV potential also depends at LO on short-range $N\!N$ interactions which makes the analysis more complicated. Considering the expected reach of future EDM experiments these results can play an important role in constraining or finding BSM physics. 

An outstanding problem are  calculations of the symmetry-breaking LECs using nonperturbative methods such as lattice QCD. Both in the PVTC and PVTV sector, very little is known about the sizes of the LECs which hampers the interpretation of the experimental data. We have discussed recent progress in calculations of the nucleon EDM arising from the $\tb$ term and quark EDMs, but stress that similar calculations for the other PVTV sources would be very important. The same can be said for calculations of the PVTC and PVTV pion-nucleon LECs.

In this work we have focused on few-nucleon systems where the scattering equations can be solved with high precision. Many experiments have been performed on heavier systems in which PVTC and PVTV effects can be significantly enhanced, see discussions in Refs.~\cite{Haxton:2013aca,Engel:2013lsa,Roberts:2014bka}. The great challenge for the future is to extend the $\chi$EFT framework discussed here beyond the few-body regime. Great progress has been made in the last couple of years in performing ab initio calculations of medium-heavy nuclei based on chiral PCTC $N\!N$ interactions \cite{Epelbaum:2012vx,Carlson:2014vla}. It would be extremely interesting to extend these calculations to include symmetry violations. As discussed above, an interesting intermediate step would be the ab initio calculation of the PVTC $\vec p \alpha$ analyzing power for which a nonzero measurement has been reported.

\subsection*{Acknowledgements}
We thank all our many collaborators for sharing their insights
into the topics discussed here.
This work is supported in part by the DFG and the NSFC
through funds provided to the Sino-German CRC 110 ``Symmetries and
the Emergence of Structure in QCD'' (Grant No. 11261130311). The work of UGM was supported in part by the Chinese Academy of Sciences CAS
              President's International Fellowship Initiative (PIFI) grant no. 2015VMA076.

\bibliographystyle{h-physrev3}
\bibliography{review}

\end{document}